\documentclass[lettersize,journal]{IEEEtran}
\usepackage{amsmath,amsfonts}
\usepackage{array}
\usepackage[caption=false,font=normalsize,labelfont=sf,textfont=sf]{subfig}
\usepackage{textcomp}
\usepackage{stfloats}
\usepackage{url}
\usepackage[hidelinks]{hyperref}
\usepackage{verbatim}
\usepackage{graphicx}
\usepackage{cite}
\usepackage{multirow}
\usepackage{xcolor}
\usepackage{algorithm}
\usepackage{algpseudocode}
\usepackage{siunitx} 

\begin{document}

\title{TDPP: Two-Dimensional Permutation-Based Protection of Memristive Deep Neural Networks}

\author{ 
    Minhui~Zou\IEEEauthorrefmark{2}\IEEEauthorrefmark{1},
    Zhenhua~Zhu\IEEEauthorrefmark{3},~
    Tzofnat~Greenberg-Toledo\IEEEauthorrefmark{1},~
    
    Orian Leitersdorf\IEEEauthorrefmark{1},~\IEEEmembership{Student Member,~IEEE,}
	Jiang~Li\IEEEauthorrefmark{1},~
    Junlong~Zhou\IEEEauthorrefmark{2},~\IEEEmembership{Member,~IEEE,}
    Yu~Wang\IEEEauthorrefmark{3},~\IEEEmembership{Fellow,~IEEE,}
    Nan~Du\IEEEauthorrefmark{4}\IEEEauthorrefmark{5},~
    and~Shahar~Kvatinsky\IEEEauthorrefmark{1},~\IEEEmembership{Senior Member,~IEEE}

    \vspace{-1ex}

	\IEEEcompsocitemizethanks{
		\IEEEcompsocthanksitem \IEEEauthorrefmark{2}School of Computer Science and Engineering, Nanjing University of Science and Technology, Jiangsu, China, 210049. \IEEEauthorrefmark{3}Department of Electrical Engineering, BNRist, Tsinghua University, Beijing, China, 100084. \IEEEauthorrefmark{4}Institute for Solid State Physics, Friedrich Schiller University Jena, Fürstengraben 1, 07743 Jena, Germany. \IEEEauthorrefmark{5}Department of Quantum Detection, Leibniz Institute of Photonic Technology (IPHT), Albert-Einstein-Str. 9, 07745 Jena, Germany. \IEEEauthorrefmark{1}Viterbi Faculty of Electrical and Computer Engineering, Technion -- Israel Institute of Technology, Haifa, Israel, 3200003.

		E-mails: minhui.zou@outlook.com, zhuzhenh18@mails.tsinghua.edu.cn, stzgrin@campus.technion.ac.il, orianl@campus.technion.ac.il, lijiang@nuaa. edu.cn, jlzhou@njust.edu.cn, yu-wang@mail.tsinghua.edu.cn, nan.du@leibniz -ipht.de, and shahar@ee.technion.ac.il.
	}
}

\markboth{Accepted to IEEE TRANSACTIONS ON COMPUTER-AIDED DESIGN OF INTEGRATED CIRCUITS AND SYSTEMS, 2023
}%
{Shell \MakeLowercase{\textit{et al.}}: A Sample Article Using IEEEtran.cls for IEEE Journals}

\IEEEpubid{0000--0000/00\$00.00~\copyright~2021 IEEE}

\maketitle

\IEEEpubid{\begin{minipage}{\textwidth}\ \\[12pt] \centering \copyright 2023 IEEE. Personal use of this material is permitted.  Permission from IEEE must be obtained for all other uses, in any current or future media, including reprinting/republishing this material for advertising or promotional purposes, creating new collective works, for resale or redistribution to servers or lists, or reuse of any copyrighted component of this work in other works. \end{minipage}} 

\begin{abstract}
    The execution of deep neural network (DNN) algorithms suffers from significant bottlenecks due to the separation of the processing and memory units in traditional computer systems.
    Emerging memristive computing systems introduce an in situ approach that overcomes this bottleneck. 
    The non-volatility of memristive devices, however, may expose the DNN weights stored in memristive crossbars to potential theft attacks.
    Therefore, this paper proposes a two-dimensional permutation-based protection (TDPP) method that thwarts such attacks.
    We first introduce the underlying concept that motivates the TDPP method: permuting both the rows and columns of the DNN weight matrices.
    This contrasts with previous methods, which focused solely on permuting a single dimension of the weight matrices, either the rows or columns.   
    While it's possible for an adversary to access the matrix values, the original arrangement of rows and columns in the matrices remains concealed. 
    As a result, the extracted DNN model from the accessed matrix values would fail to operate correctly.
    We consider two different memristive computing systems (designed for layer-by-layer and layer-parallel processing, respectively) and demonstrate the design of the TDPP method that could be embedded into the two systems.
    Finally, we present a security analysis.
    Our experiments demonstrate that TDPP can achieve comparable effectiveness to prior approaches, with a high level of security when appropriately parameterized. 
    In addition, TDPP is more scalable than previous methods and results in reduced area and power overheads.
    The area and power are reduced by, respectively, 1218$\times$ and 2815$\times$ for the layer-by-layer system and by 178$\times$ and 203$\times$ for the layer-parallel system compared to prior works. 
\end{abstract}


\section{introduction}
	Artificial intelligence (AI) techniques have enabled machines to surpass human capabilities in research areas such as image recognition and have become an integral part of society.
	AI uses advanced deep neural network (DNN) algorithms such as convolutional neural networks to accomplish its tasks~\cite{parhi2020brain}.
	The separation of processing and memory units in modern computer architecture, however, means that a tremendous amount of energy is utilized when executing the data-intensive DNN algorithms \cite{mehonic2022brain}.
	Emerging memristive computing systems have demonstrated great potential in boosting the energy efficiency of the DNN algorithms \cite{shafiee_isaac_2016,mehonic2022brain}.
	Their advantage is their ability to store the DNN weights and process them in memory, thereby avoiding the tremendous data movement between the computing and memory units \cite{mehonic2022brain}.
	
	\IEEEpubidadjcol

	\begin{figure}[t!]
		\centering
		\includegraphics[width=0.478\textwidth]{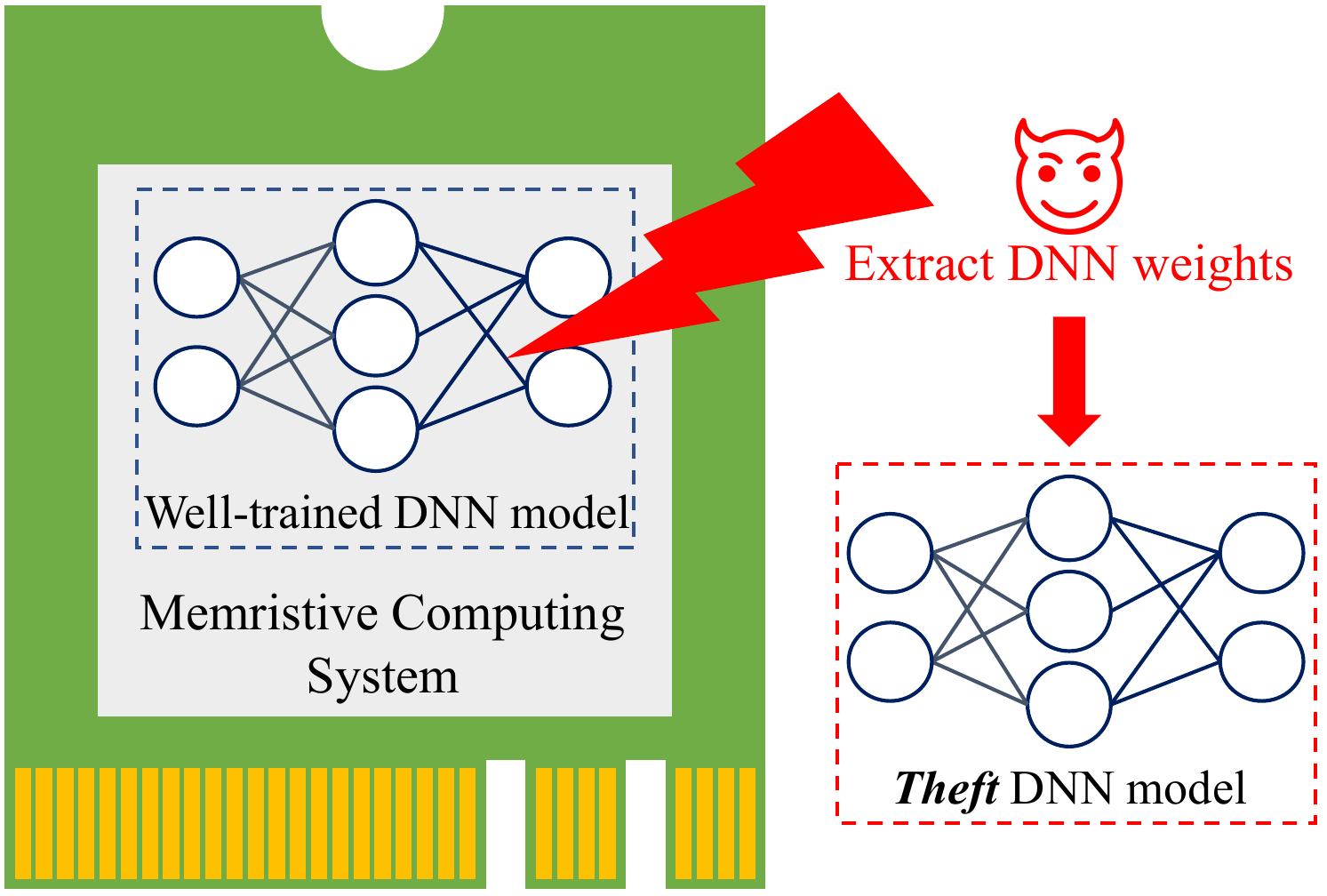}
	    \vspace{-1ex}	
        \caption{The DNN models loaded in memristive computing systems face potential theft attacks due to the non-volatility of memristor devices.}
        \label{fig:basic_idea}
        \vspace{-0ex}
	\end{figure}
	
	Despite this appealing advantage, the security of memristive computing systems has yet to receive sufficient attention.
    That is, as shown in Fig. \ref{fig:basic_idea}, DNN models stored in the memristive computing systems face theft attacks because of the non-volatility of memristive devices.
    While the memristive devices’ non-volatility might be appealing, it facilitates data theft attacks, which are real threats \cite{chhabra_i-nvmm_2011,young2015deuce,awad2016silent,zuo2019supermem} in scenarios of using memristive devices as main memory.
	If a memristor-based Dual In-line Memory Module (DIMM) is stolen, an adversary can stream out the data stored in the memory from the DIMM.
	For memristive computing systems, the current commercial memristive chips are embedded in boards with M.2 \cite{mythic_mm1076_nodate} or PCIe \cite{mythic_mp10304_nodate} interfaces.
	Moreover, the memristive chips may also be equipped with I/Os ports such as GPIOs and I$^2$C \cite{mythic_m1076_nodate}.
	These universal interfaces and ports allow an adversary to steal the data from the memristive chips.
    Thus, the adversary, having physical access to the memristive computing systems, could steal the DNN weights stored in the memristive crossbars by exploiting the data persistence of memristive devices.
    Once in possession of the DNN weights, the adversary may reverse-engineer the well-trained DNN models stored in the memristive computing systems.
    The stolen DNN models could be sold illegally to customers, resulting in copyright infringement and economic losses to the DNN model designers.
    Additionally, if the models are trained with proprietary datasets, the stolen models could leak private information, such as patients' information in a medical system, as the case may be.

	\IEEEpubidadjcol

	The existing protection methods for memristive main memory, such as counter mode encryption \cite{chhabra_i-nvmm_2011,young2015deuce,awad2016silent,zuo2019supermem} are based on encrypting the data with conventional cryptographic algorithms and decrypting them while they are being used.
	The methods, however, are not suitable for memristive computing systems because they require frequent writing operations to the memristive devices, which leads to extra high costs in both energy and latency.
	Given that the endurance property of the state-of-the-art memristive devices is limited \cite{lanza_standards_2021}, the extra writing operations could also shorten the lifetime of the memristive computing systems.
	Even worse, these methods would open an attack window for the adversary to exploit when the DNN weights on the memristive crossbars executing the DNN algorithms are decrypted.
	Though the time window may be narrow, the adversary could use side-channel analysis to pinpoint the exact execution time of each DNN layer and then turn off the systems to stream out the DNN weights of those layers.
	For instance, \cite{cai_enabling_2019} encrypted only part of the DNN weights to reduce decryption time.
	Nevertheless, this partial encryption method still involves frequent writing operations to some memristive devices, and the attack windows, though minor, persist.
    
    Another type of protection method calls for transforming the DNN weight matrices.
    It does not rely on encrypting the DNN weights; thus, the shortcomings of the above methods are avoided.
    This type of method provides round-the-clock security for the DNN weights, i.e., whenever the adversary carries out theft attacks, the DNN weights are always protected.
	\cite{zou2022enhancing} suggested selectively encoding some columns of weights as their ones' complement and leaving the others untouched.
	The adversary does not know which columns of weights are encoded, so the actual representation of the weights is hidden.
	This method, however, may increase the output value range at bitlines (BLs) and thus require a higher-precision analog-to-digital converter (ADCs) \cite{shafiee_isaac_2016}.
	Another sort of weight matrix transforming is matrix row/column permutation.
    The protection proposed by \cite{zou_security_2020} was to hide the row connections between crossbar pairs.
    Conversely, \cite{wang_low_2021} suggested grouping memristive crossbars into multiple virtual operation units (VOU) and permuting the VOUs along the column dimension.     
	Nevertheless, the existing matrix row/column permutation methods have some shortcomings and challenges that need to be countered:

	(1) \textbf{Scalability}. 
	Both methods assume the crossbar digital-to-analog converters (DACs) and ADCs are shared among wordlines (WLs) and BLs, respectively, and that they can reduce the hardware overheads of their respective protection methods by exploiting DAC/ADC multiplexing.
	Typically, for a 256 $\times$ 256 crossbar, they assume that only 16 WLs and 16 BLs are enabled simultaneously\footnote{\cite{wang_low_2021} considered 8 WLs and 8 BLs for a crossbar size of 128 $\times$ 128.}.
	In fact, the number of simultaneously activated WLs/BLs ($x$) varies, depending on the specific architecture and implementation. 
	For example, NeuRRAM \cite{wan2022compute} suggested that it is possible to activate all the crossbar rows and columns simultaneously using voltage-mode sensing instead of current-mode sensing.
	The protection method of \cite{zou_security_2020} is only applicable when $x$ is 16 because, for its protection hardware, the output of each multiplexer (MUX) in the first layer needs to be connected to all the MUXes in the middle layer.
    As to the protection method of \cite{wang_low_2021}, it is not applicable when $x$ is 1 or 256 since the crossbar row grouping mechanism is invalid, and when $x$ is large, such as 128, the method becomes insecure because the number of VOUs is minimal.

	(2) \textbf{Vulnerability}. 
    Both \cite{zou_security_2020} and \cite{wang_low_2021} only considered the security of a single protected crossbar or one crossbar pair.
    In Section \ref{sec:effectiveness_and_security}, we investigated the security aspects of the proposed TDPP in terms of the entire model, going beyond the analysis of single crossbars or crossbar pairs. 
    By adopting this broader perspective, our aim is to provide a more comprehensive understanding of the security implications associated with our approach.
    Additionally, permutation-based protection methods may be vulnerable to several types of attacks, especially divide-and-conquer attacks \cite{liu2016chosen}.
    These potential attacks, however, were not considered by them, either.
    As mentioned in above paragraph, when $x$ is large, the methods of \cite{zou_security_2020} and \cite{wang_low_2021} becomes inapplicable and insecure, respectively.

	(3) \textbf{Key strategy}.
	The protection hardware of both protection methods \cite{zou_security_2020} and \cite{wang_low_2021} is dispersed in the peripheral of every crossbar pair, complicating the peripheral design.
	Furthermore, for parallel execution of the crossbars, the protection keys also need to be near the crossbars.
	The keys would be stored in volatile memory, such as buffers or registers.
	Their papers do not clarify how the keys are generated and shared among the crossbars.

	In this paper, we propose a two-dimensional permutation-based protection (TDPP) method permuting both the rows and columns of the DNN weight matrices, which also belongs to the matrix row/column permutation class.
    TDPP differs from previous works and is more advantageous in several ways, which are summarized below:
    \begin{itemize}
        \item The TDPP method -- We offer a new method involving the permutation of both rows and columns of the weight matrices.
        Conversely, previous works exclusively addressed the permutation of a single dimension within the weight matrices, specifically either the rows or the columns.
		\item Implementation design -- We consider two different memristive computing systems (designed for layer-by-layer and layer-parallel processing, respectively) and present the design of the TDPP method for memristive computing systems that could be embedded in the two systems.
        We include the essential design parameters and key strategy.
        \item Security analysis -- We discuss the security metrics of the proposed method, including its resistance to brute-force attacks, divide-and-conquer attacks, and known-plaintext attacks. 
        Note that permutation-based protection methods do not guarantee absolute security. 
        Nevertheless, it aims to enhance security by introducing confusion and complexity to the arrangement of weight matrix rows and columns stored in memristor devices, thereby increasing the difficulty for attackers to extract correct DNN weights.
        \item Evaluation by simulation -- 
        We evaluated the maximum security provided by TDPP based on the minimal effort for divide-and-conquer attacks to succeed.
        We show that the TDPP method is highly effective, secure, and scalable. 
        It delivers up to 1218$\times$ and 2815$\times$ lower area and power, respectively, than related works \cite{zou_security_2020,wang_low_2021}.
    \end{itemize}

\section{background}
\label{sec:Threat_Model_Related_works_and_Motivation}
    \subsection{Preliminaries}
        \paragraph{Main parts of DNN algorithms}
            The main parts of DNN algorithms are convolution (Conv) and fully-connected (FC) layers.
            These algorithms are dominated by vector-matrix multiplications (VMMs) because both Conv and FC layers can be implemented with VMM operations \cite{ankit2019puma}.
            The weights of FC layers are in the form of matrices and the weights of Conv layers can also be transformed into matrices by reshaping each filter kernel into a column.
            For simplicity, we assume the weights of the Conv layers are already transformed into matrices. 
    		Thus, in this paper, \textbf{both the FC layer weights and the Conv layer weights are in the form of matrices}.
		
		\paragraph{Analogous VMMs with memristive crossbars}
    		In memristive computing systems, the memristive devices are organized in the form of crossbars.
    		When applying voltages in the WLs of memristive crossbars, the BLs of the memristive crossbars output the accumulated currents, which is analogous to VMMs. 
    		The input feature maps of DNNs are transformed into voltages by using DACs so that they can be applied to the WLs, and the accumulated current outputs at the BLs are converted back to digital values using ADCs.
    		Due to the non-negative conductance values from the memristive devices \cite{li_rram-based_2015}, a weight matrix is mapped to a pair of memristive crossbars, i.e., a positive crossbar (XB+) and a negative crossbar (XB-).
    		Additionally, because of the limited precision of memristive devices, multiple crossbar pairs are used to represent a high-precision weight matrix \cite{cai_low_2020}.
		
	\subsection{Threat Model}
    \label{sec:threat_model}
		As shown in Fig. \ref{fig:basic_idea}, the well-trained DNN models are loaded into the memristive computing systems.
		Memristive computing systems are whole chips embedded in boards with a universal interface such as M.2 or PCIe.
		We assume the adversary has physical access to the memristive computing systems \textbf{but does not own the stored DNN models and is motivated to steal them from the systems.}
        The adversary can insert the memristor-based DIMM into their own host machine, gaining access to the host memory to know the input of the first DNN layer to the memristive computing system and the output of the last DNN layer.
		We also assume the adversary can stream out the values of the memristive devices through the board interface or the I/O ports by exploiting the non-volatility of memristive devices.
		This threat model is aligned with the existing works \cite{cai_enabling_2019,zou_security_2020,wang_low_2021,zou2022enhancing}.
		The goal of the adversary is to read the DNN weights from the memristive devices.
		Once possessing the correct DNN weights, the adversary could extract the DNN models.
		Our motivation is to prevent the adversary from reading the DNN weights correctly.

\section{The TDPP Method}
\label{sec:permutation_based_encryption}
	
	\begin{figure}
		\centering
		\includegraphics[width=0.478\textwidth]{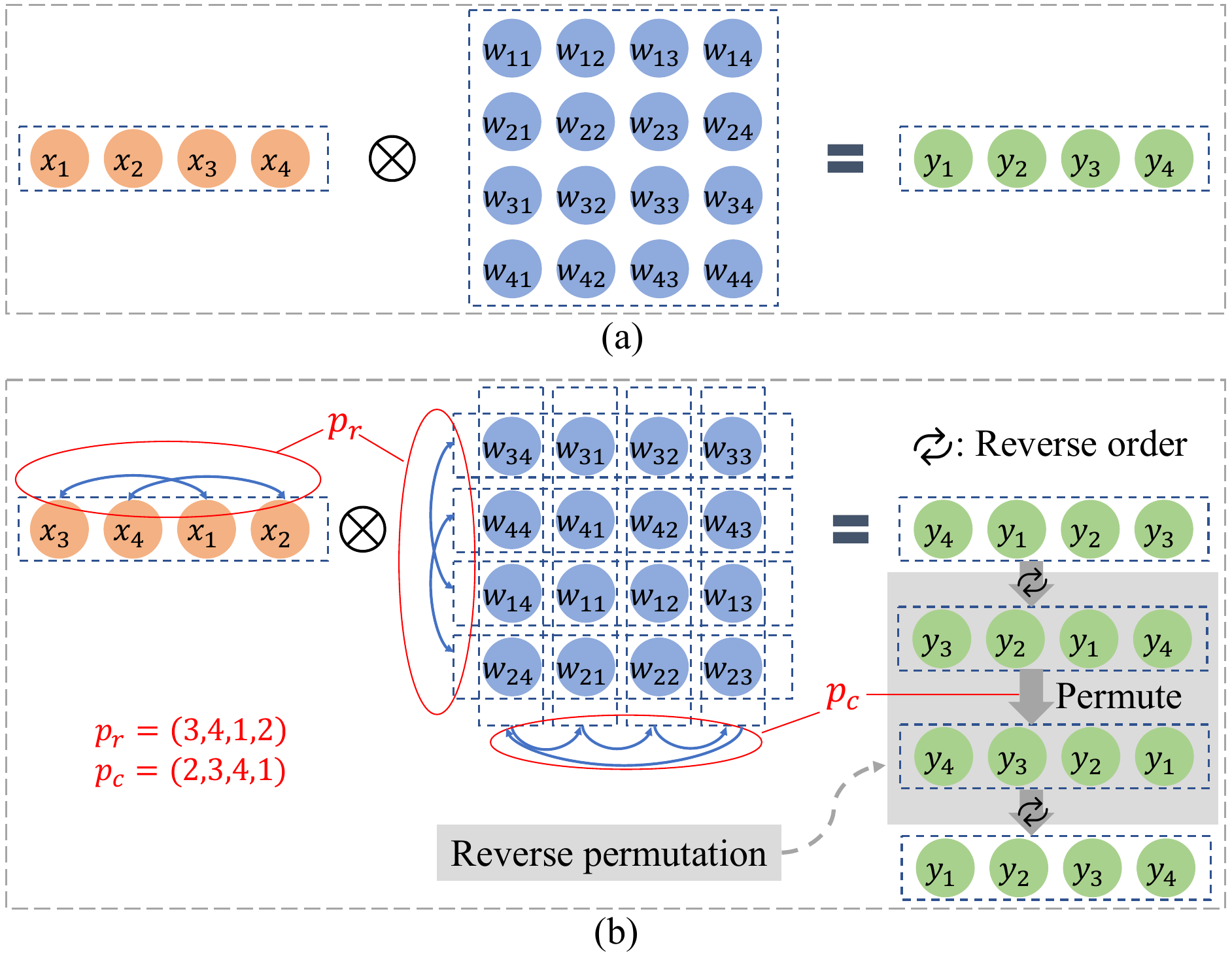}
		\caption{(a) A four-element input vector multiples an unprotected $4\times 4$ weight matrix and outputs a four-element output vector; (b) The rows and columns of the weight matrix are permuted according to $P_r$ and $P_c$, respectively; the input and output vectors need to be permuted and reverse-permuted, correspondingly, to get the correct VMM results.
        }
		\label{fig:permutation}
	\end{figure}

	Fig. \ref{fig:permutation} illustrates the basic idea of the TDPP method for protecting our weight matrix example.
    Fig. \ref{fig:permutation}(a) shows the VMM operation between a four-element input vector and a $4\times 4$ weight matrix, which is plainly mapped to the memristive devices.
	Thus, the adversary could correctly read the weight matrix values  through the corresponding memristive devices.
	Fig. \ref{fig:permutation}(b) shows the securely mapped weight matrix: the rows and columns of the original matrix have been permuted according to the vectors $P_r$ and $P_c$, respectively. 
	The vectors $P_r$ and $P_c$ indicate the permutation patterns, which are the keys. 
	For example, the vector $P_r$ being (3,4,1,2) means the 1st, 2nd, 3rd and 4th rows of the original matrix have moved to become the 3rd, 4th, 1st, and 2nd rows, respectively.
	For the correctness of the VMM operation, the input vector is also permuted according to the vector $P_r$.
	The output vector of the VMM operation between the permuted input vector and the permuted weight matrix has to be reverse-permuted to get the correct VMM result according to the vector $P_c$.
    The reverse permutation occurs by first reversing the vector, then permuting the vector, and finally reversing the vector again.
	Similarly, the weight matrix of each layer of a model is permuted independently.
	Without knowledge of $P_r$ and $P_c$, the extracted weight matrices known to the adversary are very different from the original weight matrices, so the weights of the model are well protected.

\section{TDPP Design for Memristive Computing Systems}
\label{sec:design_parameters}
    \subsection{Two Different Memristive Computing Systems}
        Fig. \ref{fig:architecture}(a) shows two memristive computing systems.
		One comprises a global arithmetic unit (AU) and a global buffer, and the other puts a tile AU and a tile buffer in each tile.
        These systems are designed for layer-by-layer and layer-parallel processing, respectively.
    	Denote them as config-1 and config-2, respectively. 
        Except for the location of the AUs and buffers, the two systems share a similar design, such as the architecture of the tiles and processing elements (PEs).
    	As shown in Fig. \ref{fig:architecture}(b), a global or tile AU consists of several digital processing modules: the adding module, the pooling module, and the activation module.
    	The system consists of many tiles for both systems, with each tile composed of multiple processing elements (PEs).
    	Each PE comprises multiple crossbar pairs.
    	The precision of both the DNN weights and the memristive devices determines the number of crossbar pairs.
    	For example, eight crossbar pairs are needed per PE when the precision of the DNN weights and memristive devices are 8 and 1, respectively.
     
	\subsection{Design of the TDPP Hardware}
    	\begin{figure*}[ht]
    		\centering
    		\includegraphics[width=1\textwidth]{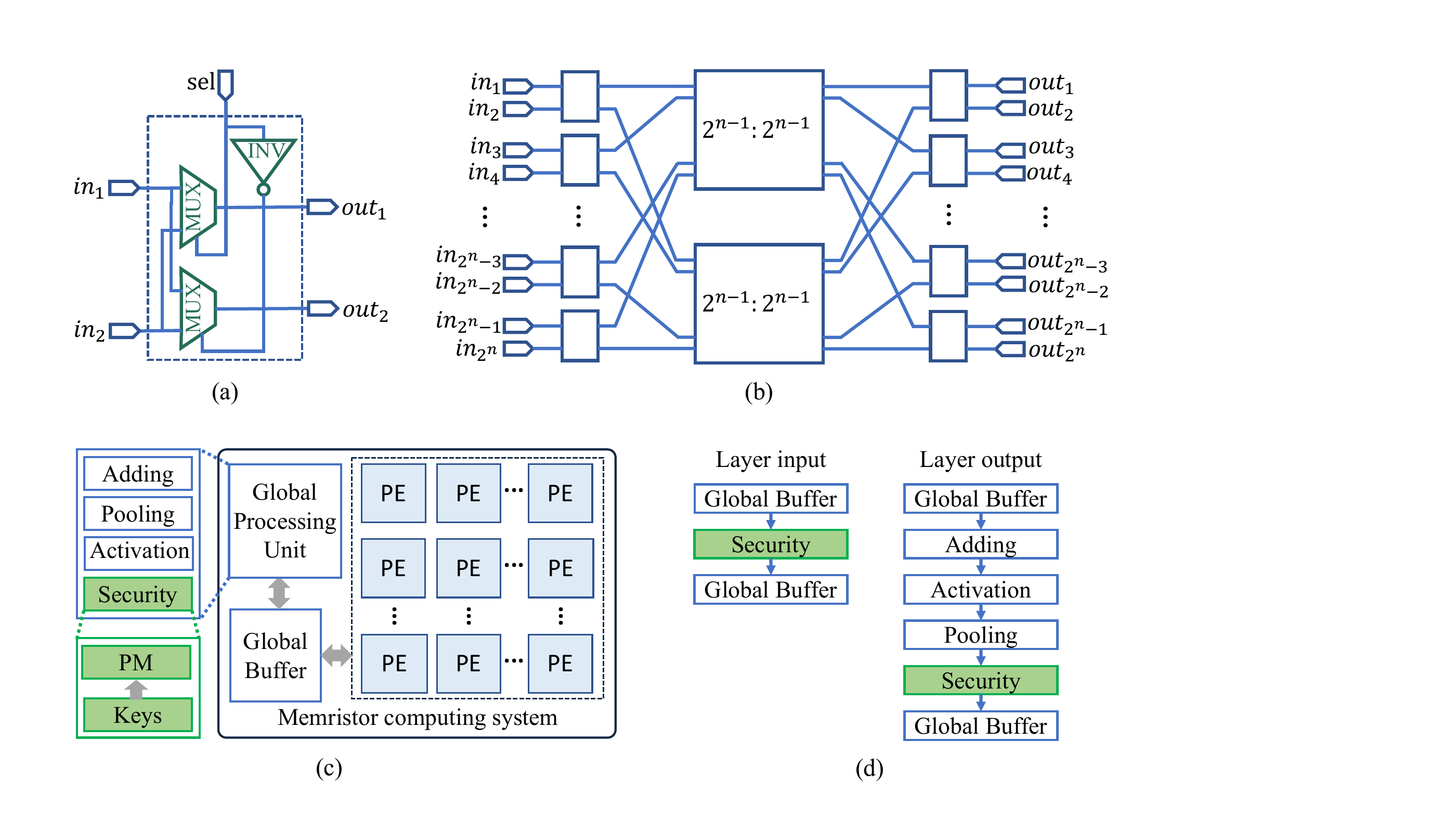}
            \vspace{-4ex}
            \caption{(a) Memristive computing systems config-1 and config-2; (b) An arithmetic unit (AU) with a TDPP hardware module embedded; (c) A 2:2 Benes Network (BN) consists of two MUXes; (d) A $2^b$:$2^b$ BN made up of two $2^{b-1}$:$2^{b-1}$ BNs and two columns of 2:2 switches \cite{huang_benes_2013}; (e) An alternative implementation of a $2^b$:$2^b$ PM (permutation module) with $k$ $2^B$:$2^B$ BNs; (f) A PM can do partial permutation.}
    		\label{fig:architecture}
    	\end{figure*}
    
		As shown in Fig. \ref{fig:architecture}(b), the TDPP design consists of a permutation module (PM), a key storage module, and a key generator.
        		
		(1) \emph{PM}: The PM is used for both permuting the layer's inputs and reverse-permuting layer's outputs.
        To minimize the hardware overhead and the system latency, we suggest implementing the PM using the Benes Network (BN)~\cite{huang_benes_2013}.
        Fig. \ref{fig:architecture}(c) shows the structure of a 2:2 BN, essentially a 2:2 switch.
        A 2:2 switch could be composed of two 2:1 MUXes.
		When the $sel$ signal is 0, the inputs $in_1$ and $in_2$ will be connected to the outputs $out_1$ and $out_2$, respectively; otherwise, the inputs will be cross-connected to the outputs.
        Fig. \ref{fig:architecture}(d) shows the structure of a $2^b$:$2^b$ BN, constructed by recursively connecting smaller-size BNs.
        Generally, a $2^b$:$2^b$ BN consists of $(2^{b-1}\times (2b-1))$ 2:2 BNs.
        Each 2:2 BN comes with a $sel$ signal, and all the signals together determine the permutation pattern.
        Denote the signals as key, the size of the key $s_{b}$ for a $2^b$:$2^b$ BN equals the number of 2:2 BNs it contains, described as
        \begin{equation} \label{eq:key_size_bn}
    		s_{b} = (2^{b-1} \times (2b-1)).
    	\end{equation}
        Two benefits are achieved when using a BN-based PM implementation.
        First, BNs are non-blocking, i.e., at any given time, all the inputs and outputs of the BNs are connected.
        The non-blocking feature is essential to avoid affecting the system throughput of the memristive computing system.
        Second, the number of 2:2 switches required by BNs is optimized, which is significant if we want to impose minimal hardware overhead on the system.
		Additionally, the vector reversing step can be done using the PM by setting the selection signals of its last $b$ columns of 2:2 switches to 1 and that of the remaining switches to 0, without additional hardware.

		We can also reduce the hardware overhead of the PM by implementing it with multiple smaller BNs instead of a big BN.
    	As shown in Fig. \ref{fig:architecture}(e), a $2^b$:$2^b$ BN could be replaced by $k$ $2^B$:$2^B$ BNs, where $k$ is the number of $2^B$:$2^B$ BNs and $2^b=k \times {2^B}$.
    	A PM consisting of $k$ $2^B$:$2^B$ BNs still simultaneously connects $2^b$ inputs and $2^b$ outputs.
    	This alternative design could reduce the hardware overhead of PMs substantially.
    	For example, a 256:256 BN could be replaced by 16 16:16 BNs to reduce the hardware overhead by approximately $53\%$.
    	Note that the hardware-reduced PM design also decreases the permutation effectiveness and security.
    	Section \ref{sec:Brute_Force_Attack} analyzes the security of the PM hardware-reduced design, and Section \ref{sec:Experiments} shows that a hardware-reduced PM design can still provide sufficient permutation and security.

		(2) \emph{Key storage}: The key storage module is an on-chip buffer comprising any volatile memory technology such as eDRAM or SRAM.
		The volatility of the key storage module ensures the keys are not accessible to the adversary when the systems are powered off.
		
\begin{figure}
			\centering
			\includegraphics[width=0.26\textwidth]{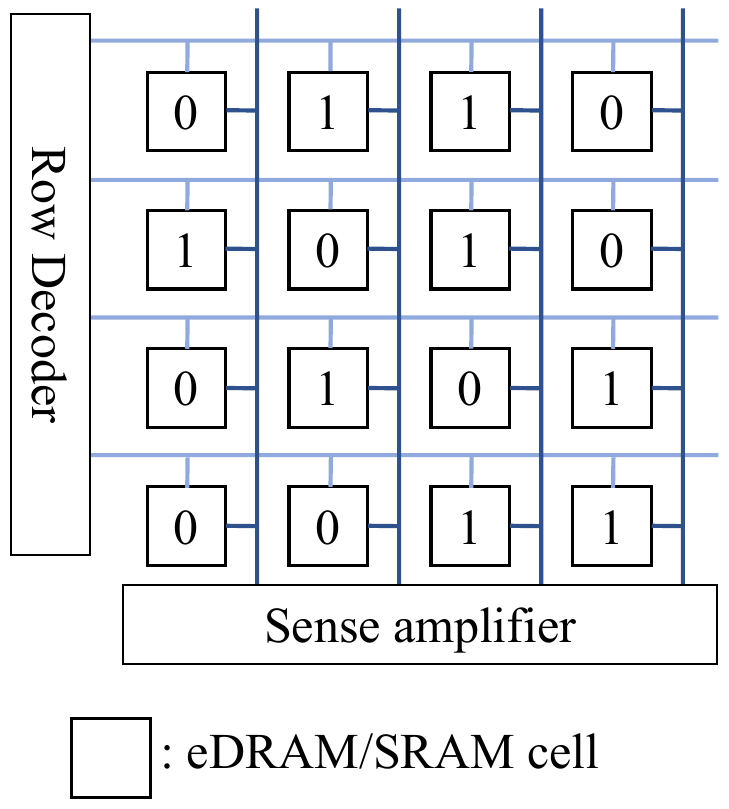}
			\caption{The key generator utilizes the startup values of eDRAM/SRAM cells, which are randomly initialized to 0 or 1 due to process variation \cite{tehranipoor2016dram,farha2020sram}.}
			\label{fig:key_generator_design}
		\end{figure}

		(3) \emph{Key generator}: The key generator generates the PM key.
		We suggest the generator be a physical unclonable function (PUF) from which the adversary cannot steal the key  \cite{mcgrath2019puf}.
        Note that the global AU and tile AUs are near the global buffer and tile buffers, respectively, and the global or tile buffer is usually a volatile eDRAM, or SRAM memory \cite{zhu_mnsim_2020}.
		We can use the global/tile buffer as a PUF by exploiting the startup values of its cells \cite{tehranipoor2016dram,farha2020sram}.
        As shown in Fig. \ref{fig:key_generator_design}, the startup values of the eDRAM/SRAM cells are randomly initialized as 0 or 1 due to process variation.
        Note that reading the startup values must be conducted before the system overrides them.
        We refer the readers to \cite{tehranipoor2016dram,farha2020sram} for detailed PUF design.

    \subsection{Embedding TDPP Hardware in Memristive Computing Systems}
    	
    	In the config-1 architecture, the DNN inference follows a layer-to-layer processing approach \cite{long2019ferroelectric,krishnan2021impact}.
    	All the layer's outputs will be transferred to the global buffer to be processed by the global AU.
    	Then, the layer's outputs will be used as inputs for the next layers and transferred to the corresponding tiles.
        We insert the TDPP hardware into the global AU.

    	In the config-2 architecture, since each tile is equipped with an AU, the output of a layer can be transferred directly to other tiles where the DNN weights of its next layer are located \cite{shafiee_isaac_2016,zhu_mnsim_2020,wan2022compute}.
    	This architecture aims at layer-parallel processing to maximize the crossbar throughput.
    	In this case, we insert a TDPP hardware module in the AU of each tile, and the key generator utilizes the cell startup values of the tile buffer.
		Note that the cell startup values of each tile buffer are different, and the key for the PM module in each tile is, therefore, unique.
    	Inserting TDPP hardware in the AU of each tile will increase the hardware overhead.
		We can, however, use the hardware-reduced PM implementation, as explained earlier, to reduce the hardware overhead.

	\subsection{Key Strategy}
		For the TDPP method described in Section \ref{sec:permutation_based_encryption}, the permutation size for a weight matrix is the same as the original weight matrix.
		The layer size of some DNN models, however, may be enormous, and its corresponding PM -- of the same size -- could be infeasible when the hardware overhead is constrained.
		To circumvent this problem, we could design a feasible-size PM and permute the rows and columns of the large weight matrices part by part separately.
		Note that in memristive computing systems, if the height or width of a layer's weight matrix is greater than that of the memristive crossbars, the matrix is divided into multiple submatrices to fit the size of the memristive crossbars.
		Each submatrix is mapped to a PE, and the PEs execute VMM operations in parallel \cite{zhu_mnsim_2020}.
        Denote the size of the memristive crossbars as $C \times C$.
        Hence, to be aligned with the crossbar parallelism, the size of the PM must be no less than the memristive crossbars, i.e., $2^b$ is at least $C$.
        For simplicity and ease of discussion, we set $2^b$ equals to $C$.
		Assume the size of a weight matrix is $m\times n$, divided into multiple submatrices by the size of crossbars.
        The rows and columns of each submatrix will be permuted independently before being mapped to the PE crossbars.  
        Note for small-size memristive crossbars, setting $2^b$ equal to $C$ might compromise security.
        For example, when $C$ is $16$, according to (\ref{eq:key_size_bn}), the key size for a 16:16 BN is only 56, which might not provide enough sufficient permutation and security.
        To address this issue, however, we can set $2^b$ as a multiple of 16, for example, 256.
        In this case, the rows and columns of every 256 submatrices will be permuted independently before being mapped to the PE crossbars.
		
        For a small weight matrix, when its height $m$ or width $n$ is less than $C$, it is possible to pad it by programming the unused memristive cells with camouflage values to increase security~\cite{zou_security_2020, wang_low_2021}.
		Usually, however, the unused cells are set into a high resistance state (HRS), and the corresponding WLs/BLs are turned off during computing to reduce the sneak paths \cite{wan2022compute}.
		Since padding small weight matrices could introduce sneaking noise, we leave them in HRS in our design.
		For the weight matrix of a DNN layer, high level security can be achieved when each submatrix is permuted with a different key.
		The resultant key storage, however, could be overwhelming.
		For example, assume a weight matrix of $4094 \times 4096$ in size and $C$ equals 256.
		The matrix is divided into 256 submatrices by every 256 rows and columns, and the rows (columns) of each submatrix are permuted with a different key using a 256:256 BN-based PM.
		According to (\ref{eq:key_size_bn}), permuting 256 rows (columns) requires a 1920-bit key.
		The required key for the whole weight matrix would be $1920 \times 256$ bits and only for permuting the weight matrix's rows or columns.
		We could reuse the key inside each weight matrix to compromise between maintaining a sufficiently high level of security and having reasonably sized key storage.
		This would mean that for a layer's weight matrix, all the submatrices share the same key and that the key for permuting the rows and the key for permuting the columns of a submatrix are the same.
		The keys for each layer, however, are the same or different than those of the other layers, depending on whether the architecture is config-1 or config-2.
		For config-1, the key is generated using the cell startup values of the global buffer, and all layers share the same key.
		For config-2, each tile is mapped with no more than a single DNN layer \cite{zhu_mnsim_2020}, and each tile has a tile buffer. 
        Hence each layer can have a unique key. 

	\subsection{Data flow}
		This section examines the system data flow to understand the effects of the TDPP hardware on the system in functionality and throughput.
		For both systems, only the initial input and final output of the DNN models are transferred between the host and the memristive computing system; all the intermediate layer results are stored in the on-chip global/tile buffer.
		
		For the config-1 architecture, each layer's inputs will be copied from the global buffer to the TDPP hardware for permutation and then back to the global buffer.
		The layer inputs will then be transferred to the tiles through the network-on-chip (NoC).
		The partial outputs from the involved tiles of a DNN layer are gathered in the global buffer and then accumulated, pooled, and activated in the global AU.
		The aggregated output will go through the TDPP hardware for reverse permutation as an additional procedure.
		
\begin{figure*}
			\centering
			\includegraphics[width=0.8\textwidth]{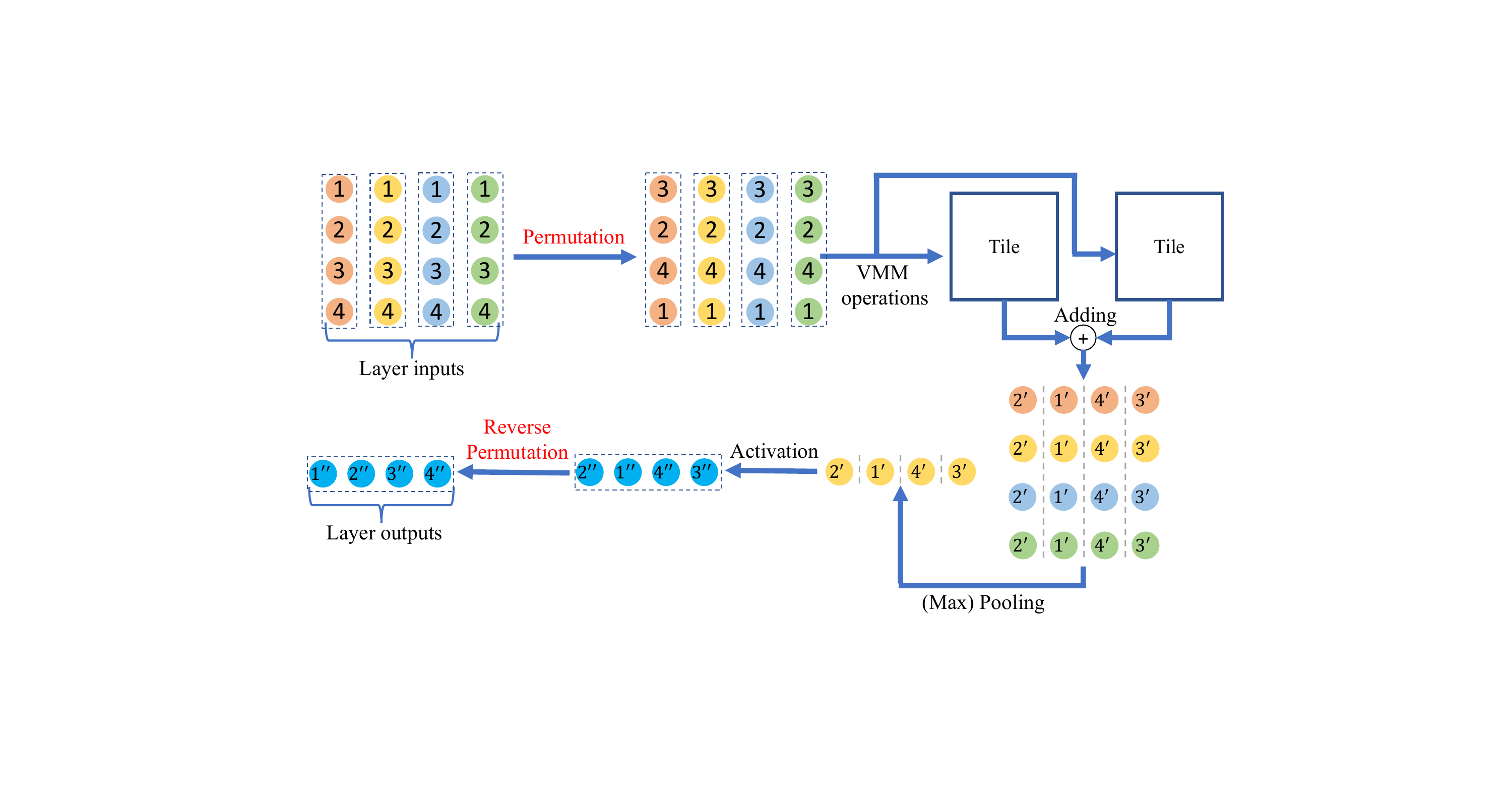}
			\caption{Example of the system data flow: \raisebox{.5pt}{\textcircled{\raisebox{-.9pt}{1}}} permuting each input vector; \raisebox{.5pt}{\textcircled{\raisebox{-.9pt}{2}}} permuted input vectors doing VMMs operations with the permuted weights loaded in tiles; \raisebox{.5pt}{\textcircled{\raisebox{-.9pt}{3}}} adding partial results from different tiles; \raisebox{.5pt}{\textcircled{\raisebox{-.9pt}{4}}} pooling aggregated outputs along the output channels (assuming the second output is the max for each output channel); \raisebox{.5pt}{\textcircled{\raisebox{-.9pt}{5}}} activating the pooled outputs; \raisebox{.5pt}{\textcircled{\raisebox{-.9pt}{6}}} reversely permuting the activated outputs to get the correct layer outputs.}
			\label{fig:dataflow}
            \vspace{-2ex}
		\end{figure*}

        Fig. \ref{fig:dataflow} illustrates a simple example of the data flow.
		Assume the input of a Conv layer is divided into four input vectors. The size of each vector is four, as the number of the input channels.
		Each input vector will be copied from the global buffer to the TDPP hardware for permutation and then copied back to the global buffer, which is ready to be transmitted to the tiles through the NoC (step \raisebox{.5pt}{\textcircled{\raisebox{-.9pt}{1}}}).
		In this case, the Conv kernels are distributed in multiple tiles.
		Each input vector will perform VMM operations with the DNN weights loaded in the tiles, and each tile will output a vector of partial results (step \raisebox{.5pt}{\textcircled{\raisebox{-.9pt}{2}}}).
		The partial results are transferred back to the global buffer and aggregated using the adding module to get an output vector of size equal to the number of output channels (four in this example) (step \raisebox{.5pt}{\textcircled{\raisebox{-.9pt}{3}}}).
		There are four input vectors, resulting in four output vectors.
		These output vectors will be pooled to become a single vector (step \raisebox{.5pt}{\textcircled{\raisebox{-.9pt}{4}}}), which then will go through the activation module (step \raisebox{.5pt}{\textcircled{\raisebox{-.9pt}{5}}}).
		The TDPP hardware will reversely permute the activated vector and, finally, copied back to the global buffer, to be ready for use as the inputs for the next layer (step \raisebox{.5pt}{\textcircled{\raisebox{-.9pt}{6}}}).
		Note that the pooling operations and activation operations are along each output channel. Therefore, the output channels are preserved.
		The reverse permutation recovers the correct order of the channels for the output vector.
		Hence, the embedded TDPP does not affect the normal functionality of memristive computing systems.
		The PM bandwidth should be at least that of the global AU or the NoC to maintain a similar system throughput.

		For the config-2 architecture, the partial VMM operation results from the involved tiles of a DNN layer gathered in one of these tiles.
		The tile AU will process the aggregated outputs and send them directly to the tiles of the next layer.
		The next layer will start processing once it gets the necessary partial outputs rather than waiting for whole outputs from the current layer \cite{zhu_mnsim_2020}.
		The proposed PM can process a partial permutation without waiting to complete a whole layer.
		As shown in Fig. \ref{fig:architecture}(f), when a PM receives a partial layer output vector, the partial vector will be padded to become the same size as a full layer output vector.
		In this case, the outputs of the first two output channels will be transferred to the next layer first. Thus the tile (knowing the key) processes VMM operations of the corresponding first and last output channels, which have top priority.
		The states of the padded elements will be set as $Z$ (high impedance state).
		After permutation, the padded elements will be discarded.
        Thus, as with the config-1 architecture, the TDPP hardware would not affect the system throughput for config-2, either.
        	
\section{Security Analysis of the TDDP Method}
\label{sec:effectiveness_and_security}

    According to the threat model outlined in Section \ref{sec:threat_model}, the adversary possesses the capability to read the values of the memristive devices, allowing them to extract the permuted weight matrices from these devices. 
    The main objective of the adversary is to reverse the permutation process and restore the rows and columns of the extracted matrices to their original arrangement, which effectively means deciphering the permutation keys generated by the key generator.
    Note that the memristive computing system is integrated into a single chip, and all components, including the PM, are on-chip. 
    Consequently, the adversary does not have control over the \textit{sel} signals of the PM. 
    Even in the scenario where the adversary gains control over the \textit{sel} signals, without knowledge of the correct permutation keys, they would be compelled to try different \textit{sel} signal combinations. 
    This process would be equivalent to attempting to restore the rows and columns of the extracted matrices to their original arrangement.
    
	\subsection{Brute-Force Attack}
	\label{sec:Brute_Force_Attack}
        Brute-force attack can be used to attempt to crack any encryption methods \cite{bernstein2005understanding}.
	    Assume a DNN model under attack has $L$ layers, and the size of the $i^{th}$ layer's weight matrix is $m^i \times n^i$ ($i \in [1,L]$).
	    According to the key strategy in Section \ref{sec:design_parameters}, the number of times a brute-force attack is undertaken $T^i_{BF}$ to recover the original weight matrix of the $i^{th}$ layer can be described as

		\begin{equation} \label{eq:brutal_force_one_layer}
			T^i_{BF} = 
			\begin{cases}
			   (B!)^k   & \text{if $m^i \ge C$ or $n^i \ge C$} 
			   \\
			   (B!)^{\lfloor{m^i/2^B}\rfloor}\cdot (m^i\%2^B)!   & \text{if $n^i \le m^i < C$}
			   \\
			   (B!)^{\lfloor{n^i/2^B}\rfloor}\cdot (n^i\%2^B)!   & \text{if $m^i < n^i < C$}
		   \end{cases}.       
	   \end{equation}

		For the config-1 architecture, as the key for each layer is the same, the effort of brute-force attacking the whole model is equal to that of attacking its biggest layer.
		Thus, the number of brute-force attacks $T_{BF}$ needed to recover all the original weight matrices of the DNN model can be described as

		\begin{equation} \label{eq:brutal_force_multi_layer_config1}
		    T_{BF} = max(T^1_{BF},T^2_{BF},...,T^L_{BF}).
		\end{equation}

		For the config-2 architecture, given that the key for each layer is different, permuting the weight matrix of each of its layers could cumulatively make the permutation space even larger.
		The model could be reverse-engineered correctly only after the weight matrices of all the layers are recovered.
		The number of brute-force attempts $T_{BF}$ needed to recover all the original weight matrices of the DNN model can be described as
		\begin{equation} \label{eq:brutal_force_multi_layer_config2}
		    T_{BF} = \prod_{i=1}^{L} {T^i_{BF}}.
		\end{equation}
        Large DNN models generally have more layers and are therefore more resistant to brute-force attacks than small DNN models.

		\subsection{Attacking Small Matrices}
		\label{sec:Attacking_small_matrices}
            Recall that for the config-1 architecture, the weight matrices of all DNN layers are permuted using the same key.
            Mapping a small matrix to a memristive crossbar, however, leaves some rows or columns in the crossbar unused, which may facilitate the adversary's brute-force attacks aiming to recover the permutation pattern for the whole DNN model.
            Furthermore, recall that for both config-1 and config-2 architecture, if the width or height of a layer’s weight matrix is larger than that of the memristor crossbars, it is divided into multiple submatrices, and the key used to permute each submatrix is the same.
            A small submatrix being mapped as a memristive crossbar may also facilitate the adversary's brute-force attacks aiming to recover the permutation pattern for the whole DNN layer.
			
\begin{figure}
				\centering
				\includegraphics[width=0.478\textwidth]{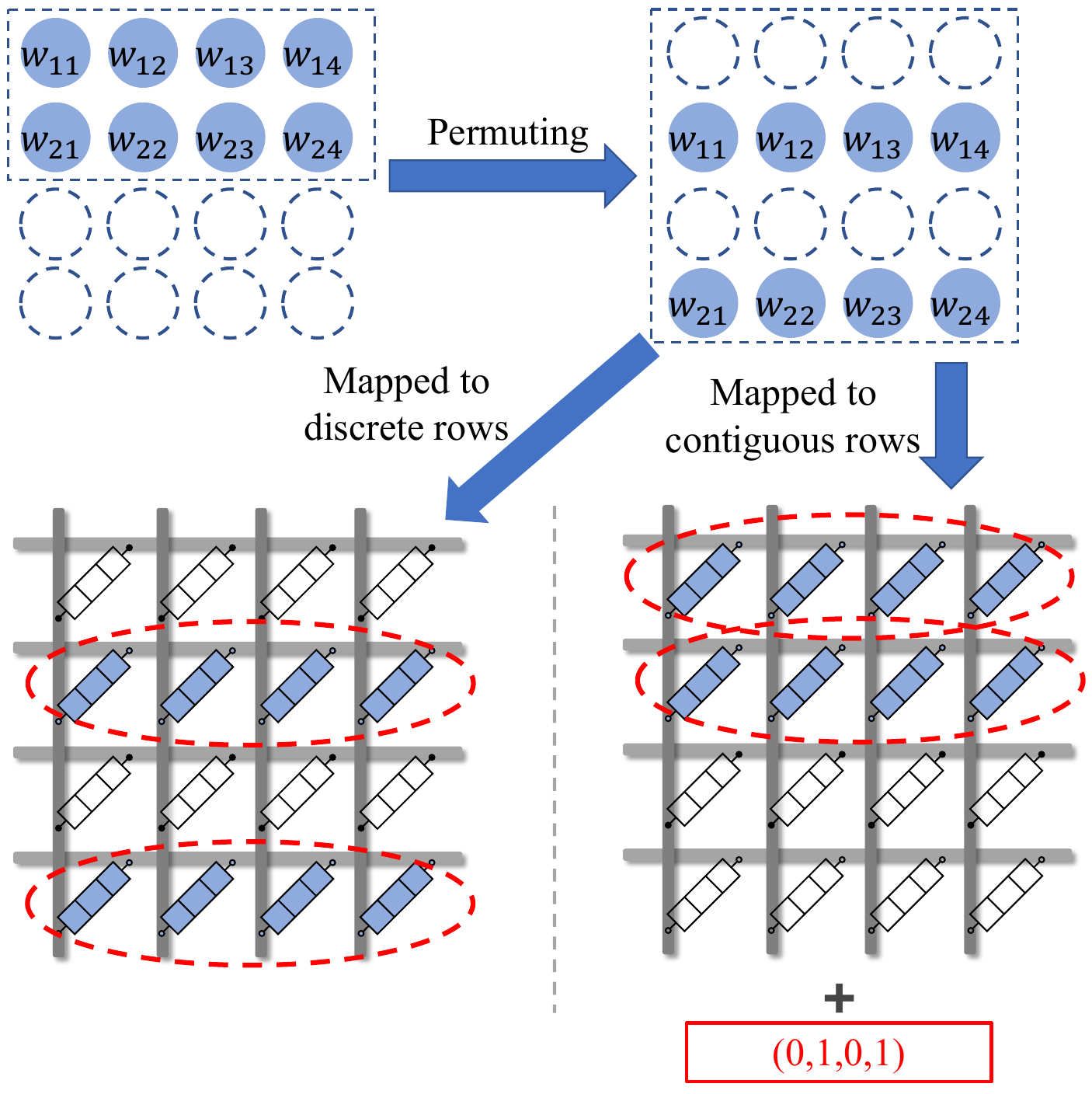}
				\caption{A matrix smaller than a memristive crossbar is permuted and mapped to (\textbf{left bottom}) discrete crossbar rows and (\textbf{right bottom}) contiguous crossbar rows with an index vector indicating the correct locations of the matrix rows.}
				\label{fig:attacking_small_matrix}
			\end{figure}

            Fig. \ref{fig:attacking_small_matrix} shows a simple example. 
			The weight matrix has two rows, $w_1$ and $w_2$, and four columns, and the crossbar size is $4 \times 4$.
			Assume the PM module is based on a 4:4 BN. Thus the number of permutation patterns is $4!$.
			After permuting the matrix, its rows become the second and fourth rows of the permuted matrix, respectively.
			If we map the permuted matrix directly to a memristive crossbar, the first and third rows of the crossbar will be left unused, from which the adversary can gain some insights into the permutation patterns.
			That is, for the permutation pattern used to permutation the example matrix, the first two permutation inputs are connected to the second and fourth permutation outputs, respectively, and the last two permutation inputs are connected to the first and third permutation outputs, respectively. 
			In this case, the possible permutation patterns are reduced from $4!$ to $2! \times 2!$, i.e., reduced by 83.33\%.
			
            To mitigate these attacks, we propose to map the rows or columns of small (sub)matrices to contiguous crossbar rows or columns, respectively, and use an \textbf{index vector} to indicate the correct location of each weight matrix row or column.
			As shown in Fig. \ref{fig:attacking_small_matrix}, $w_1$ and $w_2$ are mapped to the first and second rows of the crossbar.
			The index vector $(0,1,0,1)$ means the correct locations for $w_1$ and $w_2$ are rows two and four, respectively.
			Without knowing the index vector, the unused crossbar rows or columns do not expose information about the permutation pattern used to permute the matrix.
			The size of the index vector is equal to the number of rows or columns of the memristive crossbars.
			If the matrix size is small for both the rows and columns, we need one index vector for the rows and one for the columns.
            For TDPP, we need, at most, two index vectors for each tile.
			The index vectors are stored in the key storage of the TDPP hardware.
			Note that those vectors are generated based on the permutation keys and must not be stored in non-volatile memory.
	
	\subsection{Divide-and-Conquer Attack}
	\label{sec:Divide_and_Conquer_attack}
		Unlike conventional cryptographic algorithms, permutation-based protection methods may be vulnerable to divide-and-conquer attacks.
		For example, for a weight matrix, instead of guessing the permutation pattern for the whole matrix at once, the adversary may target only a small number of weight matrix rows or columns each time.
		The adversary expects higher and lower inference accuracy of the extracted DNN model using the correct and incorrect keys for the rows or columns, respectively.
		In this way, the original locations of the rows or columns may be recovered.
		Then the adversary will target the next set of rows or columns and continue until the locations of all the rows or columns are discovered.
		In our experiment, we use LeNet \cite{lecun_gradient-based_1998}.
		The example model consists of two Conv layers and three FC layers.
		We trained the example model with the CIFAR10 dataset \cite{krizhevsky2009learning}; the inference accuracy of the well-trained example model was $76.22\%$.
        We then permuted 25, 50, and all 75 rows of the weight matrix of the LeNet model's first layer (only permuting the rows); the inference accuracy of the extracted model was $49.4915\%$, $28.0103\%$, and $19.601\%$, respectively.
		That is, when only partial rows of the example model's first layer are permuted, the inference accuracy of the extracted model is higher than when all rows are permuted.
		
\begin{figure}
			\centering
			\includegraphics[width=0.478\textwidth]{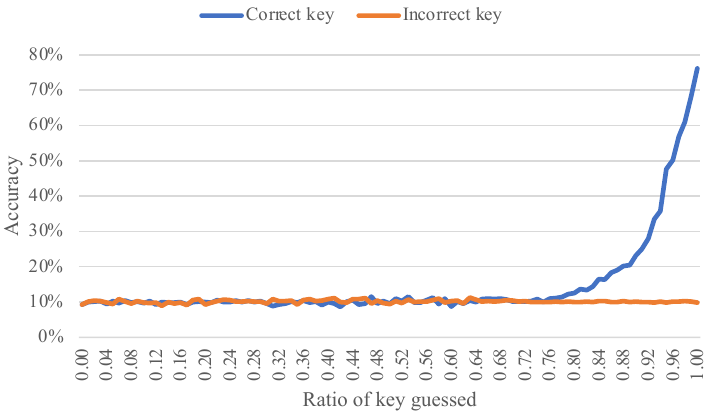}
			\caption{For config-1, the inference accuracy of the extracted example model with guessing the correct and random incorrect keys for different ratios of the key, while keeping the remaining of the key untouched.}
			\label{fig:divide_and_conquer_attack_lenet_config1}
		\end{figure}

\begin{figure}
			\centering
			\includegraphics[width=0.478\textwidth]{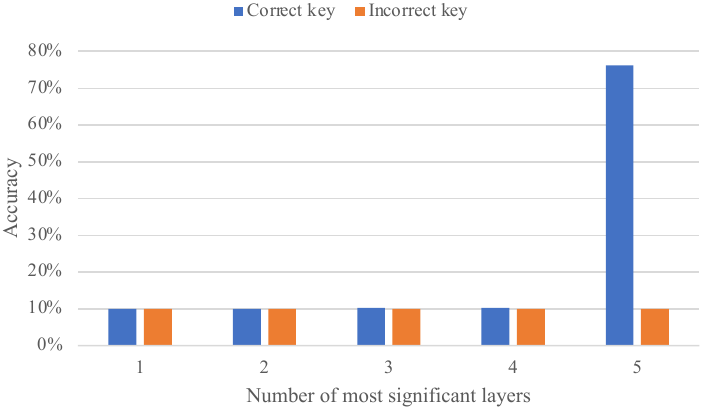}
			\caption{For config-2, the inference accuracy of the extracted example model with guessing the correct and random incorrect keys of different numbers of the most significant layers, while keeping the other layers of the model protected.}
			\label{fig:divide_and_conquer_attack_lenet_config2}
		\end{figure}

        This sort of attack, however, could not work against the proposed protection technique since, with the majority of the DNN weights protected, removing the protection of a small number of rows (columns) does not affect the model performance and the inference accuracy of the extracted model stays approximately 10\% (for the CIFAR10 dataset). 
		For config-1, we examined the inference accuracy of the extracted model by choosing the different ratios of key guess as 0.01 to 1 with 0.01 steps of the permutation key of the example model.
		For config-2, we examined the inference accuracy of the extracted model by guessing the permutation keys of the 1, 2, 3, 4, and 5 most significant layers.
		The model layer significance is measured by running the model inference with only a single layer protected.
		The lower the inference accuracy, the more significant the layer is.
		The PMs for both systems are based on a 256:256 BN.
		We compared the results of inputting correct key(s) and incorrect key(s), respectively, while keeping the other part of the model protected.
		Each experiment was carried out 40 times, and the average results were determined.
		The results are shown in Fig. \ref{fig:divide_and_conquer_attack_lenet_config1} and Fig. \ref{fig:divide_and_conquer_attack_lenet_config2}.
		For config-1, only when 74\% and above of the key is guessed correctly is the inference accuracy of the extracted model higher than that when guessing the incorrect key, i.e., the divide-and-conquer can succeed.
		For config-2, only when the keys of all layers are guessed correctly the inference accuracy of the extracted model is higher than that when guessing the incorrect key.

		We further explored the minimal effort for the divide-and-conquer attacks to succeed.
		The minimal effort is defined as the number of brute-force trial times to discover the minimal ratio of the key (for config-1) or the keys of the minimum number of DNN layers (for config-2) required to increase the inference accuracy of extracted DNN models.
		The algorithms are described as Algorithm 1 and Algorithm 2.
		For config-1, Algorithm 1 takes the brute-force attack effort $T_{BF}$ as an input.
		The output is the minimal effort for the divide-and-conquer attacks.
		The algorithm initializes the ratio $r$ as 0.01 and iterates until $r$ reaches 1 with steps of 0.01.
		For each iteration, it checks the attack sensitivity of $r$ of the permutation key.
		The attack sensitivity is defined as the relativity of the inference accuracy of the extracted DNN model when guessing correctly and incorrectly, respectively, for $r$ of the permutation key while keeping the remaining of the key untouched.
		When the correct-key results show relevant lopsidedness (at least 5\% higher) than the incorrect-key results, we regard the $r$ of the key as attack sensitive and otherwise attack insensitive.
		
        For config-2, before the algorithm, we sort the model layer significance list in descending order and store the corresponding layer indexes into a list $list1$.
		Algorithm 2 takes $list1$ and the brute-force attack effort for each layer $T^i_{BF}$ as inputs.
		The output is the minimal effort for the divide-and-conquer attacks.
		Firstly, we create a new list $list2$ to store the index of the candidate layer set that is attack sensitive.
		Similarly, the attack sensitivity is defined as the relativity of the inference accuracy of the extracted DNN model when guessing the respective correct and incorrect keys for the layer set while keeping the other layers protected.
		The algorithm keeps checking the attack sensitivity of the accumulating layer set $list2$ until $list2$ is attack-sensitive or all the layers are in $list2$.  

		\begin{algorithm}
			\label{alg:minimal_effort_config1}
			\caption{Compute the minimal effort for the divide-and-conquer attacks for config-1}
			\begin{algorithmic}[1]
				\State{Inputs: the brute force attack effort $T_{BF}$;}
				\State{Outputs: minimal effort for the divide-and-conquer attacks;}
				\For{$r=0.01; r \le 1; r=r+0.01$}
					\If{$r$ of key are attack sensitive}
						\State Break
					\EndIf
				\EndFor
			\State return $r \cdot T_{BF}$
			\end{algorithmic}
		\end{algorithm}
		
		\begin{algorithm}
			\label{alg:minimal_effort_config2}
			\caption{Compute the minimal effort for the divide-and-conquer attacks for config-2}
			\begin{algorithmic}[1]
				\State{Inputs: descending sorted model significance $list1$;}
				\State{Inputs: brute force attack effort for each layer $T^i_{BF}$, where $i \in [1,L]$ ;}
				\State{Outputs: minimal effort for the divide-and-conquer attacks;}
				\State $list2 = \{\}$;
				\While{$list1~!= NULL $ }
				 	\State $index$ = pop($list1$);
					\State $list2$.append($index$);
					\If{list2 is attack sensitive}
						\State Break
					\EndIf
				\EndWhile
			\State return $\prod_{i=1}^{L} {T^{list2[i]}_{BF}}$
			\end{algorithmic}
		\end{algorithm}

	\subsection{Known-Plaintext Attacks} 

		If the adversary knows the inputs and outputs of the PMs, then the permutation keys can easily be discovered.
        For both systems, the adversary has access to the host memory to know the input of the first DNN layer to the memristive computing system and the output of the last DNN layer.
        The input of the first DNN layer to the memristive computing system and the output of the last DNN layer, however, are irrelevant to the permutation keys.
        Only the intermediate results, permuted input vectors, and VMM operation results before reversed permutation are relevant to the permutation keys.
		The intermediate results, importantly, are stored in on-chip buffers.
		For the config-1 architecture, we assume the global buffer is implemented using eDRAM or SRAM embedded on the chip.
		For the config-2 architecture, all the tile buffers are on-chip.  
		Thus, the adversary cannot directly access the intermediate results.
		Indirectly accessing those intermediate results might be possible through side-channel analysis.
		Side-channel analysis against the intermediate layer results could be thwarted by countermeasures such as inserting fake cycles or adding noise \cite{randolph_power_2020}, and those countermeasures could be combined with TDPP to counter side-channel attacks against the intermediate results.
        
        Another potential attack involves writing specific-pattern weight matrices to the memristor crossbars. 
        In this scenario, the adversary may offload a customized DNN model with identity matrices as weight matrices to the memristor system. 
        Consequently, by processing these customized DNN weights on the memristor crossbars, the input of the first DNN layer, and the output of the last DNN layer, the difficulty of inferring intermediate results could be reduced.
        To mitigate this attack, a predefined user key can be utilized to encrypt the keys generated by the key generator within the TDPP module. 
        Instead of directly using the keys from the generator, they are XORed with the user key to create the permutation keys. 
        This way, without the correct user key, the permutation keys remain hidden from the adversary.
        This additional layer of encryption enhances the security of the system.

\section{Evaluation}
\label{sec:Experiments}
    In this section, we present our evaluation of the proposed TDPP method in terms of protection effectiveness, security, hardware area and power overheads.
	We tested the proposed method on four DNN models: AlexNet, VGG16, ResNet18, and GoogleNet.
	All the models were modified and trained on the CIFAR10 dataset, and all the models' weights were quantized as 8-bit.
	The original accuracy of the unprotected models is 86.58\%, 91.21\%, 93.27\%, and 79.88\%, respectively. 
	We ignored the errors of mapping DNN weights to the memristive devices.
	For comparison, we implemented the protection methods of \cite{zou_security_2020} and \cite{wang_low_2021} on the same models.
    We assume both methods apply a different key for each PE, that all crossbar pairs share the key inside a PE, and that the keys of each layer are different from that of other layers.
	The evaluation configuration is listed in Table \ref{tab:simulation_configuration}.
	The choices of $p$, $x$, $T$, and $B$ are \{1, 2, 4, 8\}, \{1, 2, 4, 8, 16, 32, 64, 128, 256\}, \{20, 40, 60, 80, 100\}, and \{2, 4, 8, 16, 32, 64, 128, 256\}, respectively.
	The area of the memristive cells is taken from \cite{xu_first_2020}.
	The protection modules of all the protection methods were evaluated based on 32nm CMOS technology.
	For simplicity, we assumed all the inputs, outputs, and intermediate results were 8-bit.
 	Each experiment was performed 40 times, and the average results were determined.

	\begin{table}
		\centering
		\caption{Evaluation configuration.}
		\label{tab:simulation_configuration}
		\scalebox{1}{
			\begin{tabular}{|c|c|c|c|c|c|c|c|c|c|}
				\hline
				Memristive cell (1T1R) size           &\SI{0.029}{{\micro\metre}^2} \cite{xu_first_2020} \\ \hline
				Memristive device precision           &$p$ bit                                           \\ \hline
				Crossbar size                         &$256 \times 256$                                  \\ \hline
				Number of activated WLs/BLs per cycle &$x$                                               \\ \hline
				Number of PEs per tile                &$8$                                               \\ \hline
                Number of tiles                       &$T$                                               \\ \hline
				BN size                               &$B$:$B$                                           \\ \hline
				CMOS process node                     &32nm                                              \\ \hline
	        \end{tabular}
		}
	\end{table}

	\subsection{Protection Effectiveness}
		The protection effectiveness of a protection method is defined as the inference accuracy of the protected DNN models directly extracted by the adversary.
		The lower the accuracy is, the better the effectiveness of the method. 
		The CIFAR10 dataset is 10-class; thus, when an extracted model's inference accuracy is $10\%$, the model function is randomly guessing, which is useless.
		Table \ref{tab:effectiveness} lists the effectiveness of TDPP for different values of $B$.
		For config-1, when $B$ is above 4, the extracted DNN models are nearly useless. When  $B$ is larger, the average inference accuracy shows less fluctuation, i.e., the model functions strictly as random guessing.
		For config-2, the inference accuracy of the extracted DNN models is 10\% for any value of $B$ without fluctuation.
		Based on the results of Table \ref{tab:effectiveness}, we claim that even when $B$ is very small, e.g., 4, TDPP remains highly effective for all models for both systems. 
		Moreover, the protection effectiveness is unrelated to the parameters $p$ and $x$ because TDPP is at the layer level, and the inputs/outputs of TDPP's hardware are not affected by $p$ or $x$.

		\begin{table}
			\centering
			\caption{Protection effectiveness of TDPP for different $B$.}
			\label{tab:effectiveness}
			\scalebox{0.87}{
				\begin{tabular}{|c|c|c|c|c|c|c|c|c|c|}
					\hline
					                                       &AlexNet   &VGG16     &ResNet18   &GoogleNet \\ \hline
					$B$=2, config-1, any $p$, any $x$    &24.77\%   &15.94\%   &12.82\%    &9.99\%    \\ \hline
					$B$=2, config-2, any $p$, any $x$    &10.00\%   &10.00\%   &10.00\%    &10.00\%   \\ \hline
					$B$=4, config-1, any $p$, any $x$    &10.45\%   &10.02\%   &9.88\%     &10.00\%   \\ \hline
					$B$=4, config-2, any $p$, any $x$    &10.00\%   &10.00\%   &10.00\%    &10.00\%   \\ \hline
					$B$=8, config-1, any $p$, any $x$    &9.98\%    &10.03\%   &10.16\%    &10.00\%   \\ \hline
					$B$=8, config-2, any $p$, any $x$    &10.00\%   &10.00\%   &10.00\%    &10.00\%   \\ \hline
					$B$=16, config-1, any $p$, any $x$   &10.01\%   &10.00\%   &10.06\%    &10.00\%   \\ \hline
					$B$=16, config-2, any $p$, any $x$   &10.00\%   &10.00\%   &10.00\%    &10.00\%   \\ \hline
					$B$=32, config-1, any $p$, any $x$   &10.00\%   &10.00\%   &10.02\%    &10.00\%   \\ \hline
					$B$=32, config-2, any $p$, any $x$   &10.00\%   &10.00\%   &10.00\%    &10.00\%   \\ \hline
					$B$=64, config-1, any $p$, any $x$   &10.01\%   &10.00\%   &9.98\%     &10.00\%   \\ \hline
					$B$=64, config-2, any $p$, any $x$   &10.00\%   &10.00\%   &10.00\%    &10.00\%   \\ \hline
					$B$=128, config-1, any $p$, any $x$  &10.00\%   &10.00\%   &9.99\%     &10.00\%   \\ \hline
					$B$=128, config-2, any $p$, any $x$  &10.00\%   &10.00\%   &10.00\%    &10.00\%   \\ \hline
					$B$=256, config-1, any $p$, any $x$  &10.00\%   &10.00\%   &10.00\%    &10.00\%   \\ \hline
					$B$=256, config-2, any $p$, any $x$  &10.00\%   &10.00\%   &10.00\%    &10.00\%   \\ \hline
				\end{tabular}
			}
		\end{table}

        We also compared the protection effectiveness of \cite{zou_security_2020} and \cite{wang_low_2021}.
        The method of \cite{zou_security_2020} only applies when $x$ is 16; the method \cite{wang_low_2021} is not applicable when $x$ is 1 or 256 because the grouping strategy is invalid for both cases.
		The results show that when all layers are protected, the comparison works are also effective in protecting all the models.
	
	\subsection{Security}
		The maximum security of the protection methods was estimated as the minimal effort for divide-and-conquer attacks to succeed using Algorithms 1 and 2.
        This evaluation considers factors such as PM size, architecture (config-1 or config-2), and the specific DNN model.
		Table \ref{tab:security} lists the security of TDPP for both config-1 and config-2.
        The results are shown as logarithms in base 2.
		When $B$ is above 8 and 2 for config-1 and config-2, respectively, the minimal brute-force effort requires at least $2^{256}$ attempts.
		When $B$ increases, the minimal brute-force effort also increases significantly.
		The maximum security for config-2 is at least one order of magnitude higher than that for config-1, primarily because, in the former, each layer's key is different.
		The maximum security for config-1 could be improved to be similar to config-2 by applying a strong PUF \cite{mcgrath2019puf} as the key generator so that each layer has a different key.
		From the results, we conclude that our method is highly secure when choosing a proper $B$.

		\begin{table}
			\centering
			\caption{Security of the proposed method (logarithms in base 2) for different $B$.}
			\label{tab:security}
			\scalebox{0.88}{
				\begin{tabular}{|c|c|c|c|c|c|c|c|c|c|}
					\hline
					                                       &AlexNet    &VGG16       &ResNet18 &GoogleNet \\ \hline
					$B$=2, config-1, any $p$, any $x$     &13         &13          &13       &77        \\ \hline
					$B$=2, config-2, any $p$, any $x$	  &256        &1536        &1664     &3968      \\ \hline
					$B$=4, config-1, any $p$, any $x$	  &77         &115         &38       &230       \\ \hline
					$B$=4, config-2, any $p$, any $x$	  &2304       &4992        &4992     &14592     \\ \hline
					$B$=8, config-1, any $p$, any $x$	  &256        &384         &256      &512       \\ \hline
					$B$=8, config-2, any $p$, any $x$	  &5120       &8320        &10240    &24320     \\ \hline
					$B$=16, config-1, any $p$, any $x$    &538        &717         &538      &717       \\ \hline
					$B$=16, config-2, any $p$, any $x$    &7168       &11648       &14336    &34048     \\ \hline
					$B$=32, config-1, any $p$, any $x$    &691        &922         &806      &922       \\ \hline
					$B$=32, config-2, any $p$, any $x$    &9216       &16128       &18432    &43776     \\ \hline
					$B$=64, config-1, any $p$, any $x$    &986        &1126        &986      &1267      \\ \hline
					$B$=64, config-2, any $p$, any $x$    &11264      &19712       &22528    &53504     \\ \hline
					$B$=128, config-1, any $p$, any $x$   &1331       &1498        &1331     &1498      \\ \hline
					$B$=128, config-2, any $p$, any $x$   &13312      &23296       &26624    &63232     \\ \hline
					$B$=256, config-1, any $p$, any $x$   &1536       &1728        &1728     &1728      \\ \hline
					$B$=256, config-2, any $p$, any $x$   &15360      &26880       &30720    &72960     \\ \hline
				\end{tabular}
			}
		\end{table}

		We also compared the related works with a modified Algorithm 2.
        The original Algorithm 2 keeps checking the attack sensitivity of increasing the number of DNN layers.
        In our experiment settings, the related works apply different keys for the PEs of each DNN layer.
        Thus, the modified Algorithm 2 checks the attack sensitivity of increasing the number of PEs instead of the DNN layers.
        The comparison results are listed in Table \ref{tab:security_refs}.
        The maximum security of both \cite{zou_security_2020} and \cite{wang_low_2021} is not affected by $p$ because, in our experimental setting, all crossbar pairs inside a PE share the same key.
        For the protection method of \cite{zou_security_2020}, the maximum security is high since its permutation is applied to crossbar rows, and the weight matrices of some DNN layers of the tested models have a large number of rows, so those matrices are mapped to multiple PEs.
        Each PE that applies a different permutation key increases the maximum security significantly.
        This method, however, is only applicable when $x$ is 16 due to the implementation of its protection module.
        For the protection method of \cite{wang_low_2021}, the maximum security is also high when $x$ is up to 32.
        A small $x$ means the VOUs are small and high-multiplicity MUXes/DEMUXes are used so that the permutation space is immense.
        Nevertheless, as $x$ increases, the maximum security decreases significantly. 
        For example, when $x$ is 128, \cite{wang_low_2021} randomly divides a crossbar into two VOU groups, and each group is divided into two VOUs.
        Thus, the possible permutation patterns for a single crossbar is only ${2!}^2$, and for all models, the total maximum security it provides is at most $2^{52}$, which is insufficient.

		\begin{table}
			\centering
			\caption{maximum security of the protection methods (logarithms in base 2) of \cite{zou_security_2020} and \cite{wang_low_2021}.}
			\label{tab:security_refs}
			\scalebox{0.98}{
				\begin{tabular}{|c|c|c|c|c|c|c|c|c|c|c|}
					\hline
									                  &                         &AlexNet  &VGG16    &ResNet18   &GoogleNet  \\ \hline
					\multirow{2}{*}{$x$=1, any $p$}   &\cite{zou_security_2020} &-        &-        &-          &-          \\ \cline{2-6}
											          &\cite{wang_low_2021}     &-        &-        &-          &-          \\ \hline
					\multirow{2}{*}{$x$=2, any $p$}   &\cite{zou_security_2020} &-        &-        &-          &-          \\ \cline{2-6}
											          &\cite{wang_low_2021}     &Inf      &Inf      &Inf        &Inf        \\ \hline
					\multirow{2}{*}{$x$=4, any $p$}   &\cite{zou_security_2020} &-        &-        &-          &-          \\ \cline{2-6}
											          &\cite{wang_low_2021}     &Inf      &Inf      &Inf        &Inf        \\ \hline
					\multirow{2}{*}{$x$=8, any $p$}   &\cite{zou_security_2020} &-        &-        &-          &-          \\ \cline{2-6}
											          &\cite{wang_low_2021}     &Inf      &Inf      &Inf        &Inf        \\ \hline
					\multirow{2}{*}{$x$=16, any $p$}  &\cite{zou_security_2020} &18806    &60180    &21063      &18054      \\ \cline{2-6}
											          &\cite{wang_low_2021}     &27612    &12036    &11328      &26904      \\ \hline
					\multirow{2}{*}{$x$=32, any $p$}  &\cite{zou_security_2020} &-        &-        &-          &-          \\ \cline{2-6}
											          &\cite{wang_low_2021}     &2693     &2081     &1958       &4651       \\ \hline
					\multirow{2}{*}{$x$=64, any $p$}  &\cite{zou_security_2020} &-        &-        &-          &-          \\ \cline{2-6}
											          &\cite{wang_low_2021}     &385      &275      &275        &697        \\ \hline
					\multirow{2}{*}{$x$=128, any $p$} &\cite{zou_security_2020} &-        &-        &-          &-          \\ \cline{2-6}
											          &\cite{wang_low_2021}     &40       &26       &22         &52         \\ \hline
					\multirow{2}{*}{$x$=256, any $p$} &\cite{zou_security_2020} &-        &-        &-          &-          \\ \cline{2-6}
											          &\cite{wang_low_2021}     &-        &-        &-          &-          \\ \hline
				\end{tabular}
			}
		\end{table}

	\subsection{Hardware Overhead}
    \label{sec:hardware_overhead}
    
		In this subsection, we evaluate the hardware overheads of TDPP and the related works in terms of area and power.
		The overhead is aggregated for the protection module and key storage and does not include the key generation module.

		\subsubsection{Protection module}

            Table \ref{tab:protection_modules} summarizes the required hardware modules for different protection methods. 
            The Config-1 architecture only needs one TDPP hardware module, while the Config-2 architecture needs one TDPP hardware module in each tile. 
            The protection module required by the method proposed in \cite{zou_security_2020} comprises $2x$ $(256/x)$:1 MUXes and $x$ 1:$(256/x)$ DEMUXes. 
            In this method, each crossbar pair requires one protection module. On the other hand, the size of a VOU in the method proposed in \cite{wang_low_2021} is scaled as $x^2$, and the protection module includes one $(256/x)$:1 MUX and one 1:$(256/x)$ DEMUX. 
            Again, each crossbar pair requires one protection module. 
            However, the authors of \cite{wang_low_2021} did not consider the module's bitwidth. 
            In reality, each input/output of the MUX/DEMUX is an array of $x$ 8-bit values. 
            To ensure a fair comparison, we set the bitwidth of the MUXes and DEMUXes to $8x$ using their method.

            \begin{table}
        		\centering
        		\caption{Protection modules  of different protection methods.}
        		\label{tab:protection_modules}
        		\scalebox{1}{
        			\begin{tabular}{|c|c|c|c|c|c|c|c|c|c|}
        				\hline
        				Config-1                     &One TDPP hardware module           \\ \hline
        				Config-2                     &\begin{tabular}{@{}c@{}}one TDPP hardware module \\ per tile\end{tabular}                                            \\ \hline
        				\cite{zou_security_2020}     &\begin{tabular}{@{}c@{}}$2x$ $(256/x)$:1 MUXes and $x$ 1:$(256/x)$ DEMUXes \\ per crossbar pair\end{tabular}        \\ \hline
        				\cite{wang_low_2021}         &\begin{tabular}{@{}c@{}}one $(256/x)$:1 MUX and one 1:$(256/x)$ DEMUX \\ per crossbar pair\end{tabular}              \\ \hline
        	        \end{tabular}
        		}
                \vspace{-2ex}
        	\end{table}

		\subsubsection{Key storage}
            For the proposed method, the key storage includes both the key(s) for permutation and the index vectors.
			For \cite{zou_security_2020}, for each of the $x$ WLs, the key storage for each protection module is $x \times 3 \times log_{2} (256/x)$ bits (each MUX or DEMUX needs $log_{2} (256/x)$ bits) and so the key storage for each protection module is $x \times 3 \times log_{2} (256/x) \times (256/x)$ bits.
			For \cite{wang_low_2021}, the key storage for each protection module is ($256 \times log_{2} (256/x) + log_{2} (256/x) \times 2 \times (256/x)$) bits (row activation vectors and keys for the MUX/DEMUX for each $x$ WLs).
            To reduce the key storage overhead for \cite{zou_security_2020} and \cite{wang_low_2021}, we assume all protection modules inside each PE share the same key.
			For the parallel execution of PEs, each PE will have a corresponding key storage.
			We assume all the keys are stored in eDRAM (32nm CMOS), the area and power consumption are modeled using CACTI \cite{muralimanohar2007optimizing}.

\begin{figure}
			\centering
			\includegraphics[width=0.478\textwidth]{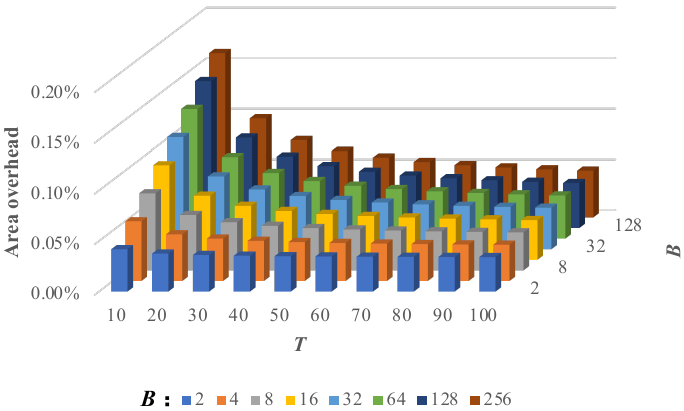}
			\caption{Area overhead of config-1 compared with memristive crossbars for $p=8$.}
            \label{fig:area_overhead_config1}
            \vspace{-2ex}
		\end{figure}

\begin{figure}
			\centering
			\includegraphics[width=0.478\textwidth]{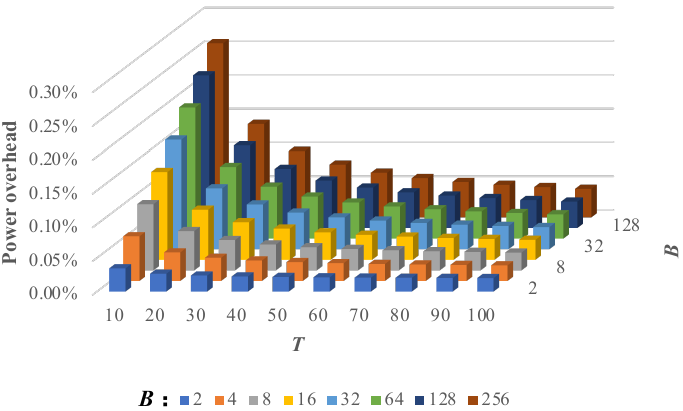}
			\caption{Power consumption overhead of config-1 compared with that of memristive crossbars for $p=8$.}
            \label{fig:power_overhead_config1}
            \vspace{-3ex}
		\end{figure}

\begin{figure}[ht]
			\centering
			\includegraphics[width=0.478\textwidth]{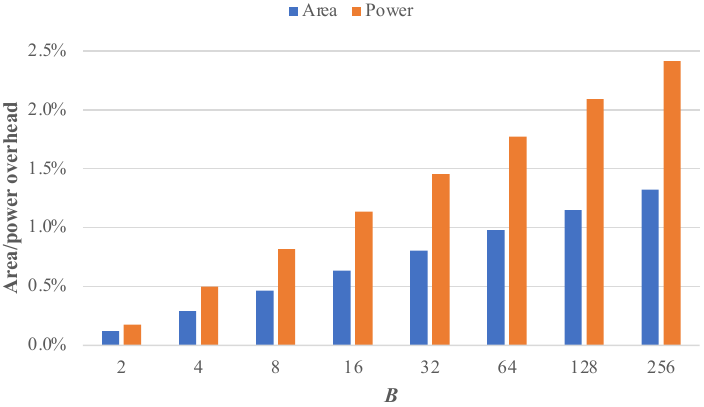}
            \vspace{-2ex}
			\caption{Area/power overhead of config-2 compared with that of memristive crossbars for $p=8$.}
			\label{fig:overhead_config2}
            \vspace{-2ex}
		\end{figure}

		Figs. \ref{fig:area_overhead_config1} and \ref{fig:power_overhead_config1} show the total area and power overheads of the proposed method for the config-1 architecture, respectively.
		When $B$ is 256, the area and power overheads are maximized, and are less than 0.16\% and 0.26\% of that of memristive crossbars for $p=8$, respectively.
		Lower $B$ would reduce the overhead.
		When $B$ is 2, compared with when $B$ is 256, the overhead could be reduced by up to approximately 74\% and 87\% for area and power, respectively.
        For the config-2 architecture, the relative overhead compared to crossbars remains constant regardless of $T$ since each tile is equipped with a protection module and a key storage module.
		Fig. \ref{fig:overhead_config2} shows the area and power overhead compared with that of crossbars for $p=8$. 

		Note that, for brevity, we only show the results for $p=8$.
		The relative overhead would decrease proportionally as $p$ decreases.
		For example, when $p$ is 1, each tile needs 8$\times$ more devices compared to $p$ = 8, and the corresponding relative overhead is one eighth of that when $p$ is 8.

        To compare with the related works, for config-1 and config-2, we set $B$ as 64 and 4, respectively, to ensure our method provide sufficient maximum security (more than $2^{986}$) for all models.		 
        Tables \ref{tab:area_overhead_comparison_normalized_p1}--\ref{tab:power_overhead_comparison_normalized_p8} list the results for different $T$, different $x$, and different $p$.
        For brevity, we only show the results for $p$ equals 1 and 8, when the gap between TDPP and the related works is the largest and smallest.
        TDPP for config-1 shows a significant advantage over that for config-2 and other protection methods mainly because it only requires one TDPP hardware module.
        The advantage increases proportionally with $T$.
        TDPP for config-1 also incurs a lower hardware overhead than the methods of \cite{zou_security_2020} and \cite{wang_low_2021} thanks to its hardware-reduced PM implementation.
        For higher $x$, the overhead advantage of TDPP versus the method of \cite{wang_low_2021} declines since a larger $x$ requires fewer MUXes and DEMUXes for the method of \cite{wang_low_2021}.
		Overall, the proposed method incur lower hardware overhead than the related works regardless of the memristive devices's precision, the number of simultaneously activated WLs/BLs, and the number of tiles. 

        It is crucial to mention that for the evaluation section, we explicitly specified the DNN weight precision as 8 bits and investigated the memristive device precision $p$ ranging from 1 bit to 8 bits.
        As for higher-precision memristive devices (such as 11-bit devices \cite{rao2023thousands}), they have the capacity to represent higher-precision DNN weights using single devices.
        Nonetheless, it is essential to reiterate that the claims of our proposed TDPP method remain valid, regardless of the memristive device precision.
        
        Furthermore, it is important to note that our primary focus has been on the implications of weight matrix permutation on the model inference accuracy, and we did not consider the non-ideality of memristor devices or the interconnect wire resistance.
        While device imperfections and interconnect wire resistance could potentially impact the model performance, it is worth noting that the security of our proposed TDPP method might be higher in such scenarios. 
        The reason for this higher security is that the minimal effort required for a divide-and-conquer attack to succeed in compromising the proposed TDPP method would likely increase rather than decrease. 
        However, evaluating the implications of device imperfections and interconnect wire resistance on our method would necessitate non-trivial and additional work. 
        As a result, we intend to address and quantify these effects in our future research.
		
		Considering that (1) TDPP achieves protection effectiveness comparable with the related works, (2) TDPP is very secure when choosing appropriate size of BN for PM implementation, and (3) with higher security ensured, TDPP imposes significantly lower area and power overheads than the related works considering different precision of memristive devices, different numbers of simultaneously activated WLs/BLs, and different number of tiles, we assert that the proposed method outperforms the related works.
		\begin{table*}
			\centering
			\caption{Normalized results of area overheads for TDPP compared to the related works when $p$ is 1.}
			\label{tab:area_overhead_comparison_normalized_p1}
			\scalebox{1}{
				\begin{tabular}{|c|c|c|c|c|c|c|c|c|c|c|}
					\hline
					                         &                         &$x$=1        &$x$=2          &$x$=4         &$x$=8         &$x$=16        &$x$=32        &$x$=64        &$x$=128       &$x$=256      \\ \hline
					\multirow{4}{*}{$T$=20}  &config-1                 &1.0$\times$  &1.0$\times$ 	 &1.0$\times$   &1.0$\times$   &1.0$\times$   &1.0$\times$   &1.0$\times$   &1.0$\times$   &1.0$\times$  \\ \cline{2-11}
										     &config-2                 &3.6$\times$  &3.6$\times$ 	 &3.6$\times$   &3.6$\times$   &3.6$\times$   &3.6$\times$   &3.6$\times$   &3.6$\times$   &3.6$\times$  \\ \cline{2-11}
					                         &\cite{zou_security_2020} &-            &-	             &-             &-             &292.2$\times$ &-             &-             &-             &-            \\ \cline{2-11}
										     &\cite{wang_low_2021}     &-            &647.1$\times$  &430.9$\times$ &287.7$\times$ &192.3$\times$ &128.5$\times$ &85.6$\times$  &56.5$\times$  &-            \\ \hline
					\multirow{4}{*}{$T$=40}  &config-1                 &1.0$\times$  &1.0$\times$ 	 &1.0$\times$   &1.0$\times$   &1.0$\times$   &1.0$\times$   &1.0$\times$   &1.0$\times$   &1.0$\times$  \\ \cline{2-11}
										     &config-2                 &5.1$\times$  &5.1$\times$ 	 &5.1$\times$   &5.1$\times$   &5.1$\times$   &5.1$\times$   &5.1$\times$   &5.1$\times$   &5.1$\times$  \\ \cline{2-11}
					                         &\cite{zou_security_2020} &-            &-	             &-             &-             &420.5$\times$ &-             &-             &-             &-            \\ \cline{2-11}
										     &\cite{wang_low_2021}     &-            &915.6$\times$  &609.6$\times$ &407.0$\times$ &272.1$\times$ &181.9$\times$ &121.1$\times$ &79.9$\times$  &-            \\ \hline
					\multirow{4}{*}{$T$=60}  &config-1                 &1.0$\times$  &1.0$\times$ 	 &1.0$\times$   &1.0$\times$   &1.0$\times$   &1.0$\times$   &1.0$\times$   &1.0$\times$   &1.0$\times$  \\ \cline{2-11}
										     &config-2                 &5.9$\times$  &5.9$\times$    &6.70$\times$  &5.9$\times$   &5.9$\times$   &5.9$\times$   &5.9$\times$   &5.9$\times$   &5.9$\times$  \\ \cline{2-11}
					                         &\cite{zou_security_2020} &-            &-	             &-             &-             &488.0$\times$ &-             &-             &-             &-            \\ \cline{2-11}
										     &\cite{wang_low_2021}     &-            &1062.6$\times$ &707.5$\times$ &472.3$\times$ &315.8$\times$ &211.0$\times$ &140.5$\times$ &92.7$\times$  &-            \\ \hline
					\multirow{4}{*}{$T$=80}  &config-1                 &1.0$\times$  &1.0$\times$ 	 &1.0$\times$   &1.0$\times$   &1.0$\times$   &1.0$\times$   &1.0$\times$   &1.0$\times$   &1.0$\times$  \\ \cline{2-11}
										     &config-2                 &6.4$\times$  &6.4$\times$ 	 &6.4$\times$   &6.4$\times$   &6.4$\times$   &6.4$\times$   &6.4$\times$   &6.4$\times$   &6.4$\times$  \\ \cline{2-11}
					                         &\cite{zou_security_2020} &-            &-	             &-             &-             &530.5$\times$ &-             &-             &-             &-            \\ \cline{2-11}
										     &\cite{wang_low_2021}     &-            &1155.3$\times$ &769.2$\times$ &513.6$\times$ &343.4$\times$ &229.5$\times$ &152.8$\times$ &100.8$\times$ &-            \\ \hline
					\multirow{4}{*}{$T$=100} &config-1                 &1.0$\times$  &1.0$\times$ 	 &1.0$\times$   &1.0$\times$   &1.0$\times$   &1.0$\times$   &1.0$\times$   &1.0$\times$   &1.0$\times$  \\ \cline{2-11}
										     &config-2                 &6.8$\times$  &6.8$\times$    &6.8$\times$   &6.8$\times$   &6.8$\times$   &6.8$\times$   &6.8$\times$   &6.8$\times$   &6.8$\times$  \\ \cline{2-11}
					                         &\cite{zou_security_2020} &-            &-	             &-             &-             &559.8$\times$ &-             &-             &-             &-            \\ \cline{2-11}
										     &\cite{wang_low_2021}     &-            &1219.1$\times$ &811.7$\times$ &541.9$\times$ &362.3$\times$ &242.1$\times$ &161.2$\times$ &106.4$\times$ &-            \\ \hline
				\end{tabular}
			}
		\end{table*}

		\begin{table*}
			\centering
			\caption{Normalized results of area overheads for TDPP compared to the related works when $p$ is 8.}
			\label{tab:area_overhead_comparison_normalized_p8}
			\scalebox{1}{
				\begin{tabular}{|c|c|c|c|c|c|c|c|c|c|c|}
					\hline
					                         &                         &$x$=1        &$x$=2          &$x$=4         &$x$=8         &$x$=16        &$x$=32        &$x$=64        &$x$=128       &$x$=256      \\ \hline
					\multirow{4}{*}{$T$=20}  &config-1                 &1.0$\times$  &1.0$\times$ 	 &1.0$\times$   &1.0$\times$   &1.0$\times$   &1.0$\times$   &1.0$\times$   &1.0$\times$   &1.0$\times$  \\ \cline{2-11}
										     &config-2                 &3.6$\times$  &3.6$\times$ 	 &3.6$\times$   &3.6$\times$   &3.6$\times$   &3.6$\times$   &3.6$\times$   &3.6$\times$   &3.6$\times$  \\ \cline{2-11}
					                         &\cite{zou_security_2020} &-            &-	             &-             &-             &54.5$\times$  &-             &-             &-             &-            \\ \cline{2-11}
										     &\cite{wang_low_2021}     &-            &101.2$\times$  &66.9$\times$  &45.0$\times$  &30.6$\times$  &20.7$\times$  &13.7$\times$  &8.5$\times$  &-            \\ \hline
					\multirow{4}{*}{$T$=40}  &config-1                 &1.0$\times$  &1.0$\times$ 	 &1.0$\times$   &1.0$\times$   &1.0$\times$   &1.0$\times$   &1.0$\times$   &1.0$\times$   &1.0$\times$  \\ \cline{2-11}
										     &config-2                 &5.1$\times$  &5.1$\times$ 	 &5.1$\times$   &5.1$\times$   &5.1$\times$   &5.1$\times$   &5.1$\times$   &5.1$\times$   &5.1$\times$  \\ \cline{2-11}
					                         &\cite{zou_security_2020} &-            &-	             &-             &-             &77.1$\times$  &-             &-             &-             &-            \\ \cline{2-11}
										     &\cite{wang_low_2021}     &-            &143.1$\times$  &94.6$\times$  &63.7$\times$  &43.2$\times$  &29.3$\times$  &19.4$\times$  &12.1$\times$  &-            \\ \hline
					\multirow{4}{*}{$T$=60}  &config-1                 &1.0$\times$  &1.0$\times$ 	 &1.0$\times$   &1.0$\times$   &1.0$\times$   &1.0$\times$   &1.0$\times$   &1.0$\times$   &1.0$\times$  \\ \cline{2-11}
										     &config-2                 &5.9$\times$  &5.9$\times$    &6.70$\times$  &5.9$\times$   &5.9$\times$   &5.9$\times$   &5.9$\times$   &5.9$\times$   &5.9$\times$  \\ \cline{2-11}
					                         &\cite{zou_security_2020} &-            &-	             &-             &-             &141.0$\times$ &-             &-             &-             &-            \\ \cline{2-11}
										     &\cite{wang_low_2021}     &-            &166.1$\times$  &109.8$\times$ &73.9$\times$  &50.2$\times$  &34.0$\times$  &22.5$\times$  &14.0$\times$  &-            \\ \hline
					\multirow{4}{*}{$T$=80}  &config-1                 &1.0$\times$  &1.0$\times$ 	 &1.0$\times$   &1.0$\times$   &1.0$\times$   &1.0$\times$   &1.0$\times$   &1.0$\times$   &1.0$\times$  \\ \cline{2-11}
										     &config-2                 &6.4$\times$  &6.4$\times$ 	 &6.4$\times$   &6.4$\times$   &6.4$\times$   &6.4$\times$   &6.4$\times$   &6.4$\times$   &6.4$\times$  \\ \cline{2-11}
					                         &\cite{zou_security_2020} &-            &-	             &-             &-             &97.3$\times$  &-             &-             &-             &-            \\ \cline{2-11}
										     &\cite{wang_low_2021}     &-            &180.6$\times$  &119.4$\times$ &80.3$\times$  &54.5$\times$  &36.9$\times$  &24.4$\times$  &15.2$\times$  &-            \\ \hline
					\multirow{4}{*}{$T$=100} &config-1                 &1.0$\times$  &1.0$\times$ 	 &1.0$\times$   &1.0$\times$   &1.0$\times$   &1.0$\times$   &1.0$\times$   &1.0$\times$   &1.0$\times$  \\ \cline{2-11}
										     &config-2                 &6.8$\times$  &6.8$\times$    &6.8$\times$   &6.8$\times$   &6.8$\times$   &6.8$\times$   &6.8$\times$   &6.8$\times$   &6.8$\times$  \\ \cline{2-11}
					                         &\cite{zou_security_2020} &-            &-	             &-             &-             &102.7$\times$ &-             &-             &-             &-            \\ \cline{2-11}
										     &\cite{wang_low_2021}     &-            &190.6$\times$  &126.0$\times$ &84.8$\times$  &57.6$\times$  &39.0$\times$  &25.8$\times$  &16.1$\times$  &-            \\ \hline
				\end{tabular}
			}
		\end{table*}

		\begin{table*}
			\centering
			\caption{Normalized results of power overheads for TDPP compared to the related works when $p$ is 1.}
			\label{tab:power_overhead_comparison_normalized_p1}
			\scalebox{1}{
				\begin{tabular}{|c|c|c|c|c|c|c|c|c|c|c|}
					\hline
					                         &                         &$x$=1        &$x$=2          &$x$=4          &$x$=8          &$x$=16         &$x$=32        &$x$=64        &$x$=128       &$x$=256      \\ \hline
					\multirow{4}{*}{$T$=20}  &config-1                 &1.0$\times$  &1.0$\times$ 	 &1.0$\times$    &1.0$\times$    &1.0$\times$    &1.0$\times$   &1.0$\times$   &1.0$\times$   &1.0$\times$  \\ \cline{2-11}
										     &config-2                 &4.7$\times$  &4.7$\times$ 	 &4.7$\times$    &4.7$\times$    &4.7$\times$    &4.7$\times$   &4.7$\times$   &4.7$\times$   &4.7$\times$  \\ \cline{2-11}
					                         &\cite{zou_security_2020} &-            &-	             &-              &-              &428.4$\times$  &-             &-             &-             &-            \\ \cline{2-11}
										     &\cite{wang_low_2021}     &-            &954.9$\times$  &636.4$\times$  &424.4$\times$  &283.2$\times$  &188.9$\times$ &125.9$\times$ &83.7$\times$  &-            \\ \hline
					\multirow{4}{*}{$T$=40}  &config-1                 &1.0$\times$  &1.0$\times$ 	 &1.0$\times$    &1.0$\times$    &1.0$\times$    &1.0$\times$   &1.0$\times$   &1.0$\times$   &1.0$\times$  \\ \cline{2-11}
										     &config-2                 &8.0$\times$  &8.0$\times$ 	 &8.0$\times$    &8.0$\times$    &8.0$\times$    &8.0$\times$   &8.0$\times$   &8.0$\times$   &8.0$\times$ \\ \cline{2-11}
					                         &\cite{zou_security_2020} &-            &-	             &-              &-              &729.9$\times$  &-             &-             &-             &-            \\ \cline{2-11}
										     &\cite{wang_low_2021}     &-            &1626.9$\times$ &1084.2$\times$ &723.1$\times$  &482.5$\times$  &321.9$\times$ &214.5$\times$ &142.6$\times$ &-            \\ \hline
					\multirow{4}{*}{$T$=60}  &config-1                 &1.0$\times$  &1.0$\times$ 	 &1.0$\times$    &1.0$\times$    &1.0$\times$    &1.0$\times$   &1.0$\times$   &1.0$\times$   &1.0$\times$  \\ \cline{2-11}
										     &config-2                 &10.4$\times$ &10.4$\times$   &10.4$\times$   &10.4$\times$   &10.4$\times$   &10.4$\times$  &10.4$\times$  &10.4$\times$  &10.4$\times$ \\ \cline{2-11}
					                         &\cite{zou_security_2020} &-            &-	             &-              &-              &953.6$\times$  &-             &-             &-             &-            \\ \cline{2-11}
										     &\cite{wang_low_2021}     &-            &2125.5$\times$ &1416.5$\times$ &944.7$\times$  &630.3$\times$  &420.5$\times$ &280.2$\times$ &186.3$\times$ &-            \\ \hline
					\multirow{4}{*}{$T$=80}  &config-1                 &1.0$\times$  &1.0$\times$ 	 &1.0$\times$    &1.0$\times$    &1.0$\times$    &1.0$\times$   &1.0$\times$   &1.0$\times$   &1.0$\times$  \\ \cline{2-11}
										     &config-2                 &12.3$\times$ &12.3$\times$ 	 &12.3$\times$   &12.3$\times$   &12.3$\times$   &12.3$\times$  &12.3$\times$  &12.3$\times$  &12.3$\times$ \\ \cline{2-11}
					                         &\cite{zou_security_2020} &-            &-	             &-              &-              &1126.2$\times$ &-             &-             &-             &-            \\ \cline{2-11}
										     &\cite{wang_low_2021}     &-            &2510.2$\times$ &1672.9$\times$ &1115.7$\times$ &744.4$\times$  &496.6$\times$ &331.0$\times$ &220.0$\times$ &-            \\ \hline
					\multirow{4}{*}{$T$=100} &config-1                 &1.0$\times$  &1.0$\times$ 	 &1.0$\times$    &1.0$\times$    &1.0$\times$    &1.0$\times$   &1.0$\times$   &1.0$\times$   &1.0$\times$  \\ \cline{2-11}
										     &config-2                 &13.8$\times$ &13.8$\times$   &13.8$\times$   &13.8$\times$   &13.8$\times$   &13.8$\times$  &13.8$\times$  &13.8$\times$  &13.8$\times$ \\ \cline{2-11}
					                         &\cite{zou_security_2020} &-            &-	             &-              &-              &1263.4$\times$ &-             &-             &-             &-            \\ \cline{2-11}
										     &\cite{wang_low_2021}     &-            &2816.0$\times$ &1876.6$\times$ &1251.6$\times$ &835.1$\times$  &557.1$\times$ &371.3$\times$ &246.8$\times$ &-            \\ \hline
				\end{tabular}
			}
		\end{table*}

		\begin{table*}
			\centering
			\caption{Normalized results of power overheads for TDPP compared to the related works when $p$ is 8.}
			\label{tab:power_overhead_comparison_normalized_p8}
			\scalebox{1}{
				\begin{tabular}{|c|c|c|c|c|c|c|c|c|c|c|}
					\hline
					                         &                         &$x$=1        &$x$=2          &$x$=4          &$x$=8          &$x$=16         &$x$=32        &$x$=64        &$x$=128       &$x$=256      \\ \hline
					\multirow{4}{*}{$T$=20}  &config-1                 &1.0$\times$  &1.0$\times$ 	 &1.0$\times$    &1.0$\times$    &1.0$\times$    &1.0$\times$   &1.0$\times$   &1.0$\times$   &1.0$\times$  \\ \cline{2-11}
										     &config-2                 &4.7$\times$  &4.7$\times$ 	 &4.7$\times$    &4.7$\times$    &4.7$\times$    &4.7$\times$   &4.7$\times$   &4.7$\times$   &4.7$\times$  \\ \cline{2-11}
					                         &\cite{zou_security_2020} &-            &-	             &-              &-              &60.9$\times$   &-             &-             &-             &-            \\ \cline{2-11}
										     &\cite{wang_low_2021}     &-            &127.9$\times$  &85.0$\times$   &56.9$\times$   &38.1$\times$   &25.6$\times$  &17.0$\times$  &11.1$\times$  &-            \\ \hline
					\multirow{4}{*}{$T$=40}  &config-1                 &1.0$\times$  &1.0$\times$ 	 &1.0$\times$    &1.0$\times$    &1.0$\times$    &1.0$\times$   &1.0$\times$   &1.0$\times$   &1.0$\times$  \\ \cline{2-11}
										     &config-2                 &8.0$\times$  &8.0$\times$ 	 &8.0$\times$    &8.0$\times$    &8.0$\times$    &8.0$\times$   &8.0$\times$   &8.0$\times$   &8.0$\times$  \\ \cline{2-11}
					                         &\cite{zou_security_2020} &-            &-	             &-              &-              &103.7$\times$  &-             &-             &-             &-            \\ \cline{2-11}
										     &\cite{wang_low_2021}     &-            &217.9$\times$  &144.9$\times$  &96.9$\times$   &65.0$\times$   &43.5$\times$  &29.0$\times$  &18.9$\times$  &-            \\ \hline
					\multirow{4}{*}{$T$=60}  &config-1                 &1.0$\times$  &1.0$\times$ 	 &1.0$\times$    &1.0$\times$    &1.0$\times$    &1.0$\times$   &1.0$\times$   &1.0$\times$   &1.0$\times$  \\ \cline{2-11}
										     &config-2                 &10.4$\times$ &10.4$\times$   &10.4$\times$   &10.4$\times$   &10.4$\times$   &10.4$\times$  &10.4$\times$  &10.4$\times$  &10.4$\times$ \\ \cline{2-11}
					                         &\cite{zou_security_2020} &-            &-	             &-              &-              &135.5$\times$  &-             &-             &-             &-            \\ \cline{2-11}
										     &\cite{wang_low_2021}     &-            &284.7$\times$  &189.3$\times$  &126.6$\times$  &84.9$\times$   &56.9$\times$  &37.8$\times$  &24.7$\times$  &-            \\ \hline
					\multirow{4}{*}{$T$=80}  &config-1                 &1.0$\times$  &1.0$\times$ 	 &1.0$\times$    &1.0$\times$    &1.0$\times$    &1.0$\times$   &1.0$\times$   &1.0$\times$   &1.0$\times$  \\ \cline{2-11}
										     &config-2                 &12.3$\times$ &12.3$\times$ 	 &12.3$\times$   &12.3$\times$   &12.3$\times$   &12.3$\times$  &12.3$\times$  &12.3$\times$  &12.3$\times$ \\ \cline{2-11}
					                         &\cite{zou_security_2020} &-            &-	             &-              &-              &160.0$\times$  &-             &-             &-             &-            \\ \cline{2-11}
										     &\cite{wang_low_2021}     &-            &336.2$\times$  &223.5$\times$  &149.5$\times$  &100.2$\times$  &67.2$\times$  &44.7$\times$  &29.1$\times$  &-            \\ \hline
					\multirow{4}{*}{$T$=100} &config-1                 &1.0$\times$  &1.0$\times$ 	 &1.0$\times$    &1.0$\times$    &1.0$\times$    &1.0$\times$   &1.0$\times$   &1.0$\times$   &1.0$\times$  \\ \cline{2-11}
										     &config-2                 &13.8$\times$ &13.8$\times$   &13.8$\times$   &13.8$\times$   &13.8$\times$   &13.8$\times$  &13.8$\times$  &13.8$\times$  &13.8$\times$ \\ \cline{2-11}
					                         &\cite{zou_security_2020} &-            &-	             &-              &-              &179.5$\times$  &-             &-             &-             &-            \\ \cline{2-11}
										     &\cite{wang_low_2021}     &-            &377.1$\times$  &250.7$\times$  &167.7$\times$  &112.5$\times$  &75.4$\times$  &50.1$\times$  &32.7$\times$  &-            \\ \hline
				\end{tabular}
			}
		\end{table*}
	
\section{Conclusion}
    The nonvolitility of memristive devices may facilitate attempts by adversaries to steal DNN weights loaded in the memristive computing systems by exploiting the data persistence.
    To mitigate this vulnerability, this paper proposed the TDPP method based on permuting both the rows and columns of the weight matrices.
    We considered two memristive computing systems and designed TDPP hardware that can be embedded in them. 
    Our experiments show that TDPP is very effective, secure, and scalable.
	Compared with similar existing works, the proposed TDPP method's area and power overhead demands are up to 1218.1$\times$ (area) and 2815.0$\times$ (power) lower and up to 178.1$\times$ (area) and 203.0$\times$ (power) lower for the two different systems, respectively.
	We also showed TDPP's security robustness against potential attacks.
	In the future, we intend to extend the proposed method to support spiking neural networks and graph neural networks.

\section*{Acknowledgments}
This paper acknowledges the funding by the German Research Foundation (DFG) Projects MemDPU (Grant Nr. DU1896/3-1), MemCrypto (Grant Nr. DU 1896/2-1), and the European Union's Horizon 2020 Research And Innovation Programme FETOpen NEU-Chip (Grant agreement No. 964877).

\bibliographystyle{IEEEtran}
\bibliography{references}

\vfill

\end{document}